\documentclass [12pt]{article}
\usepackage {color}

\definecolor{mycolor1}{rgb}{0.000,0.502,0.502}

\begin{document}
{\color{black} \textbf{Structure and magnetic order in Fe$_{{\rm 2}{\rm +} x}$V$_{{\rm 1}{\rm - 
}x}$Al}}

\bigskip

{\color{black} I Maksimov}{\color{black} $^{{\rm \ast 
}}${\color{black} , D Baabe}{\color{black} $^{{\rm 
\ast} }${\color{black} , H H Klauss}{\color{black} 
$^{{\rm \ast} }${\color{black} , F J 
Litterst}{\color{black} $^{{\rm \ast} }${\color{black} 
, R Feyerherm}{\color{black} $^{{\rm \$ 
}}${\color{black} , D M T\"{o}bbens}{\color{black} 
$^{{\rm \$} }${\color{black} , A 
Matsushita}{\color{black} $^{{\rm \S} }${\color{black} 
, S S\"{u}llow}{\color{black} $^{{\rm \ast} }$}

\bigskip

{\color{black} $^{{\rm \ast} }${\color{black} Institut 
f\"{u}r Metallphysik und Nukleare Festk\"{o}rperphysik, Technische 
Universit\"{a}t Braunschweig, 38106 Braunschweig, Germany}

{\color{black} $^{{\rm \$} }${\color{black} 
Hahn-Meitner-Institut Berlin GmbH, 14109 Berlin, Germany}

{\color{black} $^{{\rm \S} }${\color{black} National 
Research Institute for Metals, Tsukuba 305-0047, Japan}

\bigskip

Report-no: CM/121514/PAP

\bigskip

Comments: accepted for publication in J. Phys.: Condens. Matter

\bigskip

{\color{black} \textbf{Abstract:}}

{\color{black} We present a detailed structural investigation via 
neutron diffraction of differently heat treated samples 
Fe}{\color{black} $_{{\rm 2}}${\color{black} VAl and 
Fe}{\color{black} $_{{\rm 2}{\rm +} {\rm 
x}}${\color{black} V}{\color{black} $_{{\rm 1}{\rm - 
}{\rm x}}${\color{black} Al. Moreover, the magnetic behaviour of 
these materials is studied by means of $\mu $SR and 
M\"{o}ssbauer-experiments. Our structural investigation indicates that 
quenched Fe}{\color{black} $_{{\rm 2}}${\color{black} 
VAl, exhibiting the previously reported ''Kondo insulating like'' behaviour, 
is off-stoichiometric (6\% ) in its}{\color{mycolor1} 
}{\color{black} Al content. Slowly cooled 
Fe}{\color{black} $_{{\rm 2}}${\color{black} VAl is 
structurally better ordered and stoichiometric, and the microscopic magnetic 
probes establish long range ferromagnetic order below 
}{\color{black} $T$}{\color{black} 
$_{C}${\color{black} = 13K, consistent with results from bulk 
experiments. The magnetic state can be modelled as being generated by 
diluted magnetic ions in a non-magnetic matrix. Quantitatively, the required 
number of magnetic ions is too large as to be explained by a model of Fe/V 
site exchange. We discuss the implications of our findings for the ground 
state properties of Fe}{\color{black} $_{{\rm 
2}}${\color{black} VAl, in particular with respect to the role of 
crystallographic disorder.}

\bigskip

{\color{black} \textbf{1. ) Introduction}}

{\color{black} Recently, the magnetic phase diagram of the 
alloying series (Fe}{\color{black} $_{{\rm 1}{\rm - 
}x}${\color{black} V}{\color{black} 
$_{x}${\color{black} )}{\color{black} $_{{\rm 
3}}${\color{black} Al has been the focus of various detailed 
studies [1,2]. In particular, Heusler-type Fe}{\color{black} 
$_{{\rm 2}}${\color{black} VAl has been reported to exhibit a 
very unusual behaviour for an intermetallic compound, namely a 
semiconductor-like resistivity close to a magnetic instability [1]. This was 
interpreted in terms of Kondo-insulating behaviour, analogous to the system 
FeSi [3,4]. In contrast, optical conductivity studies provided evidence for 
a pseudogap in the density of states of Fe}{\color{black} $_{{\rm 
2}}${\color{black} VAl of 0.1-0.2eV [5], a view supported by 
various band structure calculations [6-8]. Notably, no temperature 
dependence of the gap features has been detected in these studies, 
apparently contradicting a Kondo insulator scenario for 
Fe}{\color{black} $_{{\rm 2}}${\color{black} VAl. 
However, the pseudogap scenario itself does not account for the unusual 
resistivity of Fe}{\color{black} $_{{\rm 
2}}${\color{black} VAl, as in the absence of magnetic 
correlations it should predict no or a positive metallic magnetoresistance, 
in conflict with experimental observations [2,9]. Therefore, in Ref. 5 it 
has been speculated that the (magneto)resistivity of 
Fe}{\color{black} $_{{\rm 2}}${\color{black} VAl 
reflects a mixture of electron excitation processes over the pseudogap and 
spin dependent scattering from impurities.}

{\color{black} Independently, on basis of specific heat and 
NMR-experiments it has been demonstrated that in 
Fe}{\color{black} $_{{\rm 2}}${\color{black} VAl 
crystallographic disorder, assumed to be present in form of atomic site 
exchange between Fe and V atoms, substantially affects the ground state 
properties of this compound [10,11]. In particular, the anomalous low 
temperature specific heat has been attributed to ferromagnetic clusters with 
a density of 0.003-0.004/unit cell, consistent with the results from NMR 
experiments. These works, as well, are in broad agreement with the results 
from band structure calculations, which predict that via Fe/V site exchange 
or crystallographic superstructure formation ferromagnetic clusters or 
long-range order might be generated in Fe}{\color{black} $_{{\rm 
2}}${\color{black} VAl [6-8]. Recently, it has been claimed that 
such impurity induced ferromagnetism was observed in the related Heusler 
compound Fe}{\color{black} $_{{\rm 2}}${\color{black} 
TiSn [12].}

{\color{black} Matsushita and Yamada [13] found that the bulk 
properties of Fe}{\color{black} $_{{\rm 
2}}${\color{black} VAl exhibit a very strong dependence on the 
applied heat treatment. They demonstrated that by way of different cooling 
procedures after an annealing stage the nature of the ground state of 
Fe}{\color{black} $_{{\rm 2}}${\color{black} VAl can 
be tuned: while quenched material exhibits the previously reported 
semiconducting-like, non-magnetic behaviour [1], a sample slowly cooled down 
to room temperature after a heat treatment shows almost metallic transport 
and a ferromagnetic transition at} {\color{black} 
$T$}{\color{black} $_{C}${\color{black} = 13K. Specific 
heat measurements assert the bulk nature of the magnetic transition in the 
slowly cooled samples.} 

In this context, the question arises if any or which material -- slowly 
cooled or quenched -- {\color{black} Fe}{\color{black} 
$_{{\rm 2}}${\color{black} VAl represents the intrinsic behaviour 
of this compound. Based on the band structure, specific heat and NMR 
results, it would have to be argued that in slowly cooled material either a 
crystallographic superstructure has been formed [6] or that a larger level 
of Fe/V site exchanges is present [7,8], both which might generate 
long-range magnetic order. Metallurgically, this is counterintuitive, as the 
slow cooling procedure should allow the system a more complete relaxation 
from internal strain and non-equilibrium site occupations, compared to the 
freezing in of such forms of crystallographic disorder in the quenched 
material. Further, since ferromagnetism in slowly cooled 
Fe}{\color{black} $_{{\rm 2}}${\color{black} VAl is a 
bulk phenomenon, in the absence of a crystallographic superstructure it 
requires a drastic increase of the number of Fe/V site exchanged positions 
from the value in the quenched material, 0.003-0.004/unit cell, probably up 
to the level of the percolation limit,} {\color{black} 
$i.e.$}{\color{black} , $\sim $10-20\% , in slowly cooled samples. Such 
a strong dependence of the number of Fe/V site exchanges on the cooling 
procedure would require a critical re-examination of previous results on 
Fe}{\color{black} $_{{\rm 2}}${\color{black} VAl. 
Finally, if the resistivity, as suggested in Ref.~5, largely arises from 
spin dependent scattering, then the very strong reduction of the resistivity 
in the slowly cooled sample rather indicates a reduction of disorder 
scattering, in conflict with a view of impurity induced ferromagnetism.} 

{\color{black} Given this contradictory experimental situation we 
decided to perform a thorough structural and magnetic investigation of 
slowly cooled Fe}{\color{black} $_{{\rm 
2}}${\color{black} VAl, concentrating here on microscopic 
experimental techniques, complementary to the bulk experiments executed so 
far [13]. In particular, we performed a high resolution neutron diffraction 
study and M\"{o}ssbauer spectroscopy on slowly cooled and quenched 
Fe}{\color{black} $_{{\rm 2}}${\color{black} VAl. In 
addition, we performed similar investigations on other samples from the 
series Fe}{\color{black} $_{{\rm 2}{\rm + 
}x}${\color{black} V}{\color{black} $_{{\rm 1}{\rm - 
}x}${\color{black} Al,} {\color{black} 
$x$}{\color{black} $>$ 0, in order to obtain reference data. Further, 
we studied slowly cooled Fe}{\color{black} $_{{\rm 
2}}${\color{black} VAl by means of muon spin relaxation. We have 
chosen these techniques for the following reasons:}

{\color{black} a.) If a crystallographic superstructure is formed in 
slowly cooled Fe}{\color{black} $_{{\rm 
2}}${\color{black} VAl, it should be detectable in diffraction 
experiments. The contrast between Fe, V and Al in x-ray diffraction, because 
of similar atomic weights, is weak, and might hinder an observation of a 
superstructure. Therefore, neutron diffraction experiments are more 
suitable, as Fe, Al and V have very different neutron scattering lengths 
yielding a bright contrast (Fe: 9.45$ \cdot $10}{\color{black} 
$^{{\rm -} {\rm 1}{\rm 5}}${\color{black} m; Al: 3.45$ \cdot 
$10}{\color{black} $^{{\rm -} {\rm 1}{\rm 
5}}${\color{black} m; V: -0.38$ \cdot 
$10}{\color{black} $^{{\rm -} {\rm 1}{\rm 
5}}${\color{black} m). Further, a superstructure might give rise 
to different local environments resolvable in M\"{o}ssbauer spectra, that is 
in form of a double peak spectrum resulting from different isomer shifts at 
inequivalent lattice sites.} 

{\color{black} b.) $^{{\rm 5}{\rm 7}}${\color{black} 
Fe-M\"{o}ssbauer spectroscopy and muon spin relaxation are used to 
characterise the magnetic properties microscopically. In addition to long 
range magnetic order these local probe techniques allow to examine 
inhomogeneous magnetic states caused by crystallographic disorder. However, 
the expected size of the average magnetic moment in 
Fe}{\color{black} $_{{\rm 2}}${\color{black} VAl 
($\sim $ 0.1$\mu $}{\color{black} $_{{\rm 
B}}${\color{black} ) is at the sensitivity limit of 
M\"{o}ssbauer-spectroscopy. Therefore, muon spin relaxation has been used as 
the most sensitive microscopic method to examine static magnetic order.} 

{\color{black} c.) If magnetic order stems from a large number ($\sim 
$10\% ) of Fe/V site exchanged positions, with the bright Fe/V contrast this 
should be resolved in a neutron diffraction experiment, and possibly in 
M\"{o}ssbauer spectroscopy and muon spin relaxation studies.}

{\color{black} d.) Repeating similar experiments for a number of 
samples Fe}{\color{black} $_{{\rm 2}{\rm + 
}x}${\color{black} V}{\color{black} $_{{\rm 1}{\rm - 
}x}${\color{black} Al,} {\color{black} 
$x$}{\color{black} $>$ 0, enables us to compare the results on 
nominally ordered Fe}{\color{black} $_{{\rm 
2}}${\color{black} VAl to those deliberately containing 
crystallographic disorder, thus allowing us to assess the relevance of 
disorder for the magnetic ground state of the systems.}

{\color{black} We note that previously M\"{o}ssbauer experiments 
have been performed on Fe}{\color{black} $_{{\rm 2}{\rm + 
}x}${\color{black} V}{\color{black} $_{{\rm 1}{\rm - 
}x}${\color{black} Al, -0.1 $ \le $} {\color{black} 
$x$}{\color{black} $ \le $ 1 [14,15]. In these works broad field 
distributions in the magnetically ordered state have been observed, which 
however have been interpreted in terms of magnetic fluctuations, in variance 
with the present view of long-range magnetic order. Further, the magnetic 
phase diagram determined in Ref. [15] differs greatly from that in [1,16], 
thus adding to the inconsistencies, and Fe}{\color{black} $_{{\rm 
2}}${\color{black} VAl was not specifically considered in these 
works.}

\bigskip

{\color{black} \textbf{2.) Experiments and results}}

{\color{black} \textbf{a.) Metallurgy: 
}}{\color{black} Stoichiometric mixtures of the constituent 
elements Fe (4N), V (3N) and Al (5N) have been arc-melted in a copper 
crucible under argon (5N) atmosphere and simultaneous Ti-gettering [13]. The 
weight loss during this process was less then 0.5\% . The specimens were cut 
from the polycrystalline ingot and sealed in evacuated quartz ampoules for 
the heat treatment. Initially, the two samples Fe}{\color{black} 
$_{{\rm 2}}${\color{black} VAl were homogenised at 1273K for 24h. 
The first sample (referred to as ''q-Fe}{\color{black} $_{{\rm 
2}}${\color{black} VAl'') was quenched in water after the heat 
treatment. The second one (referred to as ''sc-Fe}{\color{black} 
$_{{\rm 2}}${\color{black} VAl'') was slowly cooled to 553K at a 
rate of --6K/h, and subsequently furnace cooled. Other samples 
Fe}{\color{black} $_{{\rm 2}{\rm +} x}${\color{black} 
V}{\color{black} $_{{\rm 1}{\rm -} x}${\color{black} 
Al (}{\color{black} $x$}{\color{black} =0.5, 0.1, 0.02, 
0.01) were cooled down to 373K at a rate of --60K/h after homogenisation at 
1273K during 15h. Scanning electron microscopy (SEM) pictures have been 
taken on the samples, indicating that the materials consist of a homogeneous 
majority phase, with small inclusions of pure Al or Al-oxides.} 

{\color{black} The samples have been characterised via 
resistivity and susceptibility, and for Fe}{\color{black} $_{{\rm 
2}{\rm +} x}${\color{black} V}{\color{black} $_{{\rm 
1}{\rm -} x}${\color{black} Al,} {\color{black} 
$x$}{\color{black} =0.5, 0.1, 0.02, 0.01 and 
q-Fe}{\color{black} $_{{\rm 2}}${\color{black} VAl 
they exhibit the previously reported behaviour [1,16], with a suppression of 
magnetic order and a tendency towards semiconductivity as 
}{\color{black} $x$}{\color{black} = 0 is approached. In 
contrast, sc-Fe}{\color{black} $_{{\rm 
2}}${\color{black} VAl has a ferromagnetically ordered ground 
state and a more metallic resistivity, as reported in Ref. [13]. The 
transition temperatures, determined from the resistivity experiments, are: 
}{\color{black} $x $}{\color{black} = 0.5: 
}{\color{black} $T$}{\color{black} 
$_{C}${\color{black} $>$ 300K; 0.1:} {\color{black} 
$T$}{\color{black} $_{C}${\color{black} = 28K, 
}{\color{black} $x$}{\color{black} = 0.02, 0.01 and 
q-Fe}{\color{black} $_{{\rm 2}}${\color{black} VAl: 
}{\color{black} $T$}{\color{black} 
$_{C}${\color{black} $<$ 2K; sc-Fe}{\color{black} 
$_{{\rm 2}}${\color{black} VAl:} {\color{black} 
$T$}{\color{black} $_{C}${\color{black} = 13K.}

\textbf{b.) Neutron scattering results:} The structure of Fe$_{{\rm 2}{\rm + 
}x}$V$_{{\rm 1}{\rm -} x}$Al, {\color{black} 0 $ \le $ 
}{\color{black} $x$}{\color{black} $ \le $ 1}, derives 
from the cubic \textit{DO}$_{3}$ lattice of Fe$_{{\rm 3}}$Al. In this lattice Fe 
occupies two inequivalent sites (Fig. 1a): Fe I on 
($\raise.5ex\hbox{$\scriptstyle 1$}\kern-.1em/ 
\kern-.15em\lower.25ex\hbox{$\scriptstyle 2$} $ 
$\raise.5ex\hbox{$\scriptstyle 1$}\kern-.1em/ 
\kern-.15em\lower.25ex\hbox{$\scriptstyle 2$} $ 
$\raise.5ex\hbox{$\scriptstyle 1$}\kern-.1em/ 
\kern-.15em\lower.25ex\hbox{$\scriptstyle 2$} $), and Fe II on 
($\raise.5ex\hbox{$\scriptstyle 1$}\kern-.1em/ 
\kern-.15em\lower.25ex\hbox{$\scriptstyle 4$} $ 
$\raise.5ex\hbox{$\scriptstyle 1$}\kern-.1em/ 
\kern-.15em\lower.25ex\hbox{$\scriptstyle 4$} $ 
$\raise.5ex\hbox{$\scriptstyle 1$}\kern-.1em/ 
\kern-.15em\lower.25ex\hbox{$\scriptstyle 4$} $) and 
($\raise.5ex\hbox{$\scriptstyle 3$}\kern-.1em/ 
\kern-.15em\lower.25ex\hbox{$\scriptstyle 3$} $ 
$\raise.5ex\hbox{$\scriptstyle 3$}\kern-.1em/ 
\kern-.15em\lower.25ex\hbox{$\scriptstyle 3$} $ 
$\raise.5ex\hbox{$\scriptstyle 3$}\kern-.1em/ 
\kern-.15em\lower.25ex\hbox{$\scriptstyle 3$} $), respectively. It is 
assumed that for {\color{black} $x $}{\color{black} $<$ 1} 
the V ion almost entirely replaces Fe on the Fe I site [1]. For $x$ = 1, that 
is for Fe$_{{\rm 2}}$VAl, this leads to a fully ordered Heusler lattice, 
with Fe, V and Al each occupying separate cubic sublattices. Band structure 
calculations [6-8] indicated that in Fe$_{{\rm 2}}$VAl each Fe ion occupying 
an Fe I site carries a magnetic moment. This requires a site exchange of V 
with an Fe II ion, which might cause, either via superstructure formation 
($i.e.$, V occupying exclusively one Fe II site) or a sufficiently large level of 
random Fe II/V site exchanges, a magnetically ordered state or anomalous 
behaviour from diluted magnetic clusters.

In Fig. 1b we plot the calculated neutron diffraction intensities for above 
three structural modifications of Fe$_{{\rm 2}}$VAl: the fully ordered 
Heusler lattice, 10\% random Fe II/V site exchange and a superstructure with 
V entirely occupying one Fe II site. Qualitative and quantitative 
differences between the calculated spectra are clearly visible, indicating 
that in a high resolution neutron diffraction experiment the level of Fe 
II/V site exchanged positions should be resolvable down to about 3\% Fe on V 
sites.

Powder neutron diffraction data on Fe$_{{\rm 2}{\rm +} x}$V$_{{\rm 1}{\rm - 
}x}$Al, {\color{black} 0 $ \le $} {\color{black} 
$x$}{\color{black} $ \le $ 0.5}, have been taken on the 
Fine-Resolution-Powder-Diffractometer E9 of the Hahn-Meitner-Institute (HMI) 
in Berlin [17]. The diffractograms have been recorded in the region 2$\Theta 
$ = 5 - 155\r{} , with an incident neutron wavelength $\lambda $ = 
1.7964(1){\AA} at temperatures $T$ = 50K. Full Rietveld structure refinements 
of the diffraction data were performed employing the program 
WinPLOTR/FULLPROF [18]. Typical results are presented in Fig. 2, where we 
plot the data for the sample with nominal composition Fe$_{{\rm 2}{\rm 
.}{\rm 1}}$V$_{{\rm 0}{\rm .}{\rm 9}}$Al. In the plot we include the refined 
fit, Bragg peak positions and the difference between fit and data. Spectra 
of similar high statistics have been taken for both samples (q- and sc-) 
Fe$_{{\rm 2}}$VAl, and for Fe$_{{\rm 2}{\rm +} x}$V$_{{\rm 1}{\rm -} x}$Al, 
$x$ = 0.02, 0.5. 

All spectra correspond to the fully ordered cubic \textit{Fm3m} Heusler structure with 
some site disorder; a formation of a crystallographic superstructure has not 
been detected for any sample. In addition, for all samples between one and 
five Bragg peaks from an impurity phase, aside from those reflections coming 
from the \textit{Fm3m} lattice, have been observed. In view of the SEM results, the 
second phase probably is pure Al or Al-oxide; the relative intensity of the 
corresponding Bragg peaks is low (between 0.3 to 1.3\% for the largest 
peak), implying a small volume amount of this second phase (about 1\% ). We 
have tested if the refinements depend upon taking the secondary phase into 
account, but did not find a significant influence on the results. This 
reflects that the residual value of R$_{{\rm B}{\rm r}{\rm a}{\rm g}{\rm 
g}}$ is almost completely determined by the mismatch between the fitted 
profile functions and the experimental data for the Bragg peaks of the cubic 
\textit{Fm3m} phase.

More specifically, to refine our data we used as input the cubic Heusler 
lattice, but modified it to incorporate different types of disorder: a.) 
assuming nominal composition; b.) allowing for off-stoichiometry of Fe, V 
and Al; c.) considering possible site exchange between Fe, V and Al. 
Consistently, for all samples except for Fe$_{{\rm 2}{\rm .}{\rm 
5}}$V$_{{\rm 0}{\rm .}{\rm 5}}$Al the closest agreement between refinement 
and experimental data has been observed for models assuming Al 
off-stoichmetry. For Fe$_{{\rm 2}{\rm .}{\rm 5}}$V$_{{\rm 0}{\rm .}{\rm 
5}}$Al the models b.) and c.) did not yield a significantly improved 
solution compared to a.). Since Fe$_{{\rm 2}{\rm .}{\rm 5}}$V$_{{\rm 0}{\rm 
.}{\rm 5}}$Al is ferromagnetically ordered ($\mu _{{\rm o}{\rm r}{\rm d}}$ 
= 0.85$\mu _{{\rm B}}$), magnetic scattering contributes to the spectrum. 
Yet, including a ferromagnetic scattering contribution in the refinement did 
only marginally improve the fit. We stress that we could not find for any of 
our samples evidence for a significant level ($>$ 3\% ) of Fe II/V site 
exchange. 

In Table I we summarise the results of our refinements of the neutron 
diffraction data on Fe$_{{\rm 2}{\rm +} x}$V$_{{\rm 1}{\rm - 
}x}$Al{\color{black} , 0 $ \le $} {\color{black} 
$x$}{\color{black} $ \le $ 0.5}. Overall, the refinements yield very 
good agreement with the experimental data, with R$_{{\rm B}{\rm r}{\rm 
a}{\rm g}{\rm g}}$ values {\color{black} $\sim $2\%}  for 
{\color{black} 0 $ \le $} {\color{black} 
$x$}{\color{black} $ \le $ 0.1}. The isotropic thermal displacement 
parameters B$_{{\rm i}{\rm s}{\rm o}}$, which for V we coupled to that of Al 
because of the small neutron scattering length of V, exhibit hardly any $x$ 
dependence. In particular, no significant difference of the B$_{{\rm i}{\rm 
s}{\rm o}}$ parameters for sc- and q-Fe$_{{\rm 2}}$VAl has been detected. 
However, lattice parameters and the measured compositions both show 
significant differences between the two samples Fe$_{{\rm 2}}$VAl. 

Previously, it has been reported that the lattice parameters of Fe$_{{\rm 
2}{\rm +} x}$V$_{{\rm 1}{\rm -} x}$Al are decreasing for decreasing $x$ [1]. 
Specifically, while for $x$ varying from 1 to 0.5 a linear variation of the 
lattice parameter with $x$ has been found, subsequently a levelling off occurs 
as $x$ approaches 0, until finally for $x$ < 0 the lattice expands with $x$. This 
result is echoed in our experiments: upon reduction of $x$ from 0.5 to 0.02 we 
observe a non-linear suppression of the lattice parameter $a$ with $x$. This is 
illustrated in Fig. 3, where we plot the lattice parameters of Fe$_{{\rm 
2}{\rm +} x}$V$_{{\rm 1}{\rm -} x}$Al as function of $x$. Remarkably, at $x$ = 0, 
$i.e.$, for Fe$_{{\rm 2}}$VAl, we find two significantly different values of $a$ for 
slowly cooled and quenched material: for sc-Fe$_{{\rm 2}}$VAl it is about 
0.02\% larger than for q-Fe$_{{\rm 2}}$VAl. The relevance of this difference 
is indicated in Fig. 3: the construction with the dashed lines demonstrates 
that the value of the lattice parameter for sc-Fe$_{{\rm 2}}$VAl corresponds 
to that of Fe$_{{\rm 2}{\rm .}{\rm 0}{\rm 5}}$V$_{{\rm 0}{\rm .}{\rm 9}{\rm 
5}}$Al. According to the phase diagram from Ref. [16] Fe$_{{\rm 2}{\rm 
.}{\rm 0}{\rm 5}}$V$_{{\rm 0}{\rm .}{\rm 9}{\rm 5}}$Al is ferromagnetically 
ordered below {\color{black} $T$}{\color{black} 
$_{C}${\color{black} = 15K}. Therefore, we argue that the slow 
cooling procedure causes an expansion of the lattice in sc-Fe$_{{\rm 
2}}$VAl, thus generating negative applied pressure and the onset of 
ferromagnetic order below {\color{black} 
$T$}{\color{black} $_{C}${\color{black} = 13K.}

The lattice parameter of q-Fe$_{{\rm 2}}$VAl falls out of the range of the 
other, more slowly cooled samples. We believe that this reflects the 
significant Al off-stoichiometry of this sample. According to our data the 
actual composition is Fe$_{{\rm 2}}$VAl$_{{\rm 0}{\rm .}{\rm 9}{\rm 4}{\rm 
(}{\rm 4}{\rm )}}$, well out of the range of nominal stoichiometry. In 
contrast, for all other samples nominal and actual compositions are 
identical within experimental uncertainty. More specifically, we directly 
compare the Bragg spectra measured for sc- and q-Fe$_{{\rm 2}}$VAl, after 
normalising them for their overall intensities and correcting the latter for 
the angular shift from the difference in the lattice parameters. This 
procedure yields a matching of the positions of the Bragg peaks of sc- and 
q-Fe$_{{\rm 2}}$VAl, as demonstrated in Fig. 4 for part of the spectra. We 
then can directly determine the intensity difference between the two spectra 
I$_{{\rm s}{\rm c}{\rm -} {\rm F}{\rm e}{\rm 2}{\rm V}{\rm A}{\rm 
l}}$-I$_{{\rm q}{\rm -} {\rm F}{\rm e}{\rm 2}{\rm V}{\rm A}{\rm l}}$, which 
is included in Fig. 4. Variations of the intensities for different Bragg 
peaks are resolvable for the two spectra. We quantify the intensity 
variations by calculating the relative difference of the areas, 1 - $\Sigma 
$ (I$_{{\rm q}{\rm -} {\rm F}{\rm e}{\rm 2}{\rm V}{\rm A}{\rm l}}$ ($\Theta 
$) / I$_{{\rm s}{\rm c}{\rm -} {\rm F}{\rm e}{\rm 2}{\rm V}{\rm A}{\rm 
l}}$($\Theta $))$\Delta \Theta $ under each Bragg peak. In Fig. 5 we 
present the result of this analysis, with the relative Bragg peak 
intensities for sc-Fe$_{{\rm 2}}$VAl in the lower panel, and the intensity 
variation in the upper one. 

The \textit{Fm3m} Heusler lattice consists of four interpenetrating fcc sublattices with 
origins at $A$(0 0 0)$, B$($\raise.5ex\hbox{$\scriptstyle 1$}\kern-.1em/ 
\kern-.15em\lower.25ex\hbox{$\scriptstyle 4$} $ 
$\raise.5ex\hbox{$\scriptstyle 1$}\kern-.1em/ 
\kern-.15em\lower.25ex\hbox{$\scriptstyle 4$} $ 
$\raise.5ex\hbox{$\scriptstyle 1$}\kern-.1em/ 
\kern-.15em\lower.25ex\hbox{$\scriptstyle 4$} $)$, C$($\raise.5ex\hbox{$\scriptstyle 
1$}\kern-.1em/ \kern-.15em\lower.25ex\hbox{$\scriptstyle 2$} $ 
$\raise.5ex\hbox{$\scriptstyle 1$}\kern-.1em/ 
\kern-.15em\lower.25ex\hbox{$\scriptstyle 2$} $ 
$\raise.5ex\hbox{$\scriptstyle 1$}\kern-.1em/ 
\kern-.15em\lower.25ex\hbox{$\scriptstyle 2$} $) and 
$D$($\raise.5ex\hbox{$\scriptstyle 3$}\kern-.1em/ 
\kern-.15em\lower.25ex\hbox{$\scriptstyle 3$} $ 
$\raise.5ex\hbox{$\scriptstyle 3$}\kern-.1em/ 
\kern-.15em\lower.25ex\hbox{$\scriptstyle 3$} $ 
$\raise.5ex\hbox{$\scriptstyle 3$}\kern-.1em/ 
\kern-.15em\lower.25ex\hbox{$\scriptstyle 3$} $). Bragg reflections are 
produced by either all even or all odd Miller indices with the three 
structure amplitudes:

$F_{{\rm 1}}  \sim $ [($f_{A} - f_{C}$)$^{{\rm 2}}$+($f_{B} - f_{D}$)$^{{\rm 
2}}$]$^{{\rm \raise.5ex\hbox{$\scriptstyle 1$}\kern-.1em/ 
\kern-.15em\lower.25ex\hbox{$\scriptstyle 2$}} }$    \quad      for $h,k,l $all odd,

$F_{{\rm 2}}$\textit{} $\sim  f_{A} - f_{B} + f_{C} - f_{D}$\textit{}       \quad         for 
$\raise.5ex\hbox{$\scriptstyle 1$}\kern-.1em/ 
\kern-.15em\lower.25ex\hbox{$\scriptstyle 2$} $($h+k+l$) = 2$n $+1

$F_{{\rm 3}}$\textit{} $\sim  f_{A} + f_{B} + f_{C} + f_{D} $     \quad          for 
$\raise.5ex\hbox{$\scriptstyle 1$}\kern-.1em/ 
\kern-.15em\lower.25ex\hbox{$\scriptstyle 2$} $($h+k+l$) = 2$n$

\noindent
with $n$ an integer, and f$_{A,B,C,D}$ as average scattering factors for the 
atoms $A$,$ C $= Fe,$\quad  B $= V and $D $= Al, respectively. For identical stoichiometry of 
I$_{{\rm s}{\rm c}{\rm -} {\rm F}{\rm e}{\rm 2}{\rm V}{\rm A}{\rm l}{\rm 
}}$and I$_{{\rm q}{\rm -} {\rm F}{\rm e}{\rm 2}{\rm V}{\rm A}{\rm l}}$ no 
variation of the peak intensities $\raise.5ex\hbox{$\scriptstyle 
1$}\kern-.1em/ \kern-.15em\lower.25ex\hbox{$\scriptstyle 2$} $($h+k+l$) = 2$n$ is 
expected. In contrast, experimentally we find for these peaks on average a 
larger intensity in sc-Fe$_{{\rm 2}}$VAl than in q-Fe$_{{\rm 2}}$VAl. This 
proves that a stoichiometry difference exists between the two samples. 
Further, for $\raise.5ex\hbox{$\scriptstyle 1$}\kern-.1em/ 
\kern-.15em\lower.25ex\hbox{$\scriptstyle 2$} $($h+k+l$) = 2$n $+1 we find consistently 
that I$_{{\rm s}{\rm c}{\rm -} {\rm F}{\rm e}{\rm 2}{\rm V}{\rm A}{\rm 
l}{\rm} }$$<$ I$_{{\rm q}{\rm -} {\rm F}{\rm e}{\rm 2}{\rm V}{\rm A}{\rm l}}$, 
which in view of the different $F3$ reflects that $f_{D:{\rm q}{\rm -} {\rm 
F}{\rm e}{\rm 2}{\rm V}{\rm A}{\rm l}{\rm} }$$< $$f_{D:{\rm s}{\rm c}{\rm - 
}{\rm F}{\rm e}{\rm 2}{\rm V}{\rm A}{\rm l}}$, $i.e.$, a smaller Al concentration 
in the quenched sample compared to sc-Fe$_{{\rm 2}}$VAl. Because of the 
comparatively small intensities of peaks [$h k l$] = all odd, it yields a large 
error in the difference I$_{{\rm s}{\rm c}{\rm -} {\rm F}{\rm e}{\rm 2}{\rm 
V}{\rm A}{\rm l}}$-I$_{{\rm q}{\rm -} {\rm F}{\rm e}{\rm 2}{\rm V}{\rm 
A}{\rm l}}$, prohibiting conclusions on the structural properties from the 
intensity variations of these peaks.

Refining the neutron spectra assuming stoichiometric composition Fe:V:Al = 
2:1:1 yields for sc-Fe$_{{\rm 2}}$VAl a value R$_{{\rm B}{\rm r}{\rm a}{\rm 
g}{\rm g}}$ = 1.9 \% , which is significantly smaller than that of 
q-Fe$_{{\rm 2}}$VAl, R$_{{\rm B}{\rm r}{\rm a}{\rm g}{\rm g}}$ = 2.7~\% . 
Therefore, on basis of our neutron diffraction study of differently heat 
treated samples Fe$_{{\rm 2}{\rm +} x}$V$_{{\rm 1}{\rm -} x}$Al we conclude 
that sc-Fe$_{{\rm 2}}$VAl, which is magnetically ordered below $T_{C}$ = 13K, 
structurally is much better ordered than q-Fe$_{{\rm 2}}$VAl. Primarily, the 
disorder in q-Fe$_{{\rm 2}}$VAl arises from Al vacancies, while Fe/V site 
exchange is not observable for any sample within the resolution of our 
experiment. 

In addition to our structural investigation we attempted to study the 
magnetically ordered phases of sc-Fe$_{{\rm 2}}$VAl and Fe$_{{\rm 2}{\rm 
.}{\rm 1}}$V$_{{\rm 0}{\rm .}{\rm 9}}$Al using the focusing diffractometer 
E6 at the HMI. We were unable to resolve intensity differences between 
measurements taken above and below $T_{C}$ in our powder neutron diffraction 
experiments, indicating that for both sc-Fe$_{{\rm 2}}$VAl and Fe$_{{\rm 
2}{\rm .}{\rm 1}}$V$_{{\rm 0}{\rm .}{\rm 9}}$Al the ground states are 
ferromagnetically ordered with magnetic moments smaller than the resolution 
limit at E6 of about 0.5$\mu _{{\rm B}}$.

{\color{black} \textbf{c.) $\mu $SR-experiments: 
}}{\color{black} We performed time-differential muon spin 
relaxation experiments in zero external field (ZF-$\mu 
$SR)}{\color{black} \textbf{} }{\color{black} on above 
powdered specimen of slowly cooled sc-Fe}{\color{black} $_{{\rm 
2}}${\color{black} VAl in the temperature range 3 - 225K at the 
GPS spectrometer of the Paul Scherrer Institute, Switzerland. The sample was 
mounted with ultra thin aluminium tape in a gas flow cryostat and an 
electronic veto logic was used to register only the decay positron signals 
from muons stopped in the sample.} 

{\color{black} The muon spin polarisation as a function of time 
is reconstructed by forming the asymmetry between the numbers of positrons 
emitted parallel and antiparallel to the original muon spin polarisation 
direction. As described in [19] the experimentally determined asymmetry 
}{\color{black} $A(t)$}{\color{black} is given as 
}{\color{black} $A(t)=A$}{\color{black} $_{0 
}${\color{black} $G(t), $}{\color{black} where 
}{\color{black} $A$}{\color{black} 
$_{0}${\color{black} $\sim $ 0.21}{\color{black} $_{ 
}${\color{black} is the experimentally determined intrinsic 
asymmetry of the positron detectors and} {\color{black} $G(t) $}{\color{black} 
the normalised polarisation function of}{\color{black} \textit{} }{\color{black} 
the muon ensemble implanted in the sample.} 

{\color{black} In zero external field a possible source for a 
relaxation of the muon spin ensemble are static local magnetic fields at the 
muon site,} {\color{black} $B$}{\color{black} 
$_{loc}${\color{black} , due to finite electronic magnetic 
moments. The individual muon spin shows a Larmor precession around 
}{\color{black} $B$}{\color{black} $_{loc 
}${\color{black} at a frequency $\omega $ = $\gamma 
$}{\color{black} $_{{\rm \mu} {\rm} }${\color{black} 
$B$}{\color{black} $_{loc}$, {\color{black} with $\gamma 
$}{\color{black} $_{{\rm \mu} }$/2$\pi $ = 13.55MHz/kG. For a 
magnetically ordered polycrystalline sample an isotropic spatial averaging 
over all angles between {\color{black} 
$B$}{\color{black} $_{loc}$ and the initial muon polarisation 
$P_{\mu} $(0) yields a polarisation function {\color{black} 
$G$}{\color{black} (}{\color{black} 
$t$}{\color{black} ) = 1/3 + 2/3 cos($\omega 
$}{\color{black} $_{} t$) exp(-($\lambda _{{\rm s}{\rm t}{\rm 
a}{\rm t}}  t$)$^{{\rm \beta} }${\color{black} ) [20]. The 
relaxation of the oscillating part describes an inhomogeneous distribution 
of} {\color{black} $B$}{\color{black} $_{loc}$ for the 
total muon ensemble. If the average value of {\color{black} 
$B$}{\color{black} $_{loc}$ is zero and the form of the distribution 
is Gaussian, {\color{black} $G(t) $}{\color{black} is given 
by the Kubo-Toyabe function} {\color{black} 
$G$}{\color{black} (}{\color{black} 
$t$}{\color{black} ) = 1/3 + 2/3 (1-$\Delta 
$}{\color{black} $^{{\rm 2}}${\color{black} 
}{\color{black} $t$}{\color{black} $^{{\rm 
2}}${\color{black} ) exp(-$\raise.5ex\hbox{$\scriptstyle 
1$}\kern-.1em/ \kern-.15em\lower.25ex\hbox{$\scriptstyle 2$} 
$}{\color{black} $^{{\rm} }${\color{black} $\Delta 
$}{\color{black} $^{{\rm 2}}${\color{black} 
}{\color{black} $t$}{\color{black} $^{{\rm 
2}}${\color{black} ) [20].} {\color{black} In general, 
a static local field distribution in a polycrystalline sample will always 
lead to a longitudinal non-relaxing 1/3-tail and a transverse relaxing 
2/3-signal fraction. If the field distribution becomes dynamic with a slow 
fluctuation rate $\nu $} $<<$ ({\color{black} $\lambda 
$}{\color{black} $_{{\rm s}{\rm t}{\rm a}{\rm 
t}}${\color{black}} or {\color{black} $\Delta $),} the 
1/3-tail will exhibit a relaxation $\sim $ exp(-{\color{black} 
$\lambda $}{\color{black} $_{{\rm l}{\rm o}{\rm n}{\rm 
g}}${\color{black}} {\color{black} 
$t$}{\color{black} ). For fast fluctuations ($\nu $ 
$>>$}{\color{black} \textit{} }{\color{black} ($\lambda 
$}{\color{black} $_{{\rm s}{\rm t}{\rm a}{\rm 
t}}${\color{black} or $\Delta $)) a single component relaxation 
function} {\color{black} $G$}{\color{black} 
(}{\color{black} $t$}{\color{black} ) = exp(-($\lambda 
$}{\color{black} $_{{\rm d}{\rm y}{\rm 
n}}${\color{black}} {\color{black} 
$t$}{\color{black} )}{\color{black} $^{{\rm \beta 
}}${\color{black} ) can be employed}{\color{black} 
$.$}

{\color{black} Typical ZF-$\mu $SR spectra of 
Fe}{\color{black} $_{{\rm 2}}${\color{black} VAl are 
depicted in Fig. 6. At temperatures below 13K a two component structure of a 
quasi-static spectrum exists: a fast relaxing transversal part of nearly 2/3 
total signal amplitude and a slowly relaxing longitudinal part. No 
oscillating signal fraction is detected. Between 13 and 40K the fast 
relaxing signal fraction gradually disappears and above 40K only a single, 
slowly relaxing signal is observed.} 

{\color{black} The data sets can be described by a 
phenomenological polarisation function of the form}

{\color{black} $A$}{\color{black} 
(}{\color{black} $t$}{\color{black} )= 
}{\color{black} $A$}{\color{black} $_{0}$ (F$_{{\rm 
s}{\rm l}{\rm o}{\rm w}}$ exp(-({\color{black} $\lambda 
$}{\color{black} $_{{\rm l}{\rm o}{\rm n}{\rm 
g}}${\color{black}} {\color{black} 
$t$}{\color{black} )}{\color{black} $^{{\rm \alpha 
}}${\color{black} ) + (1- F}{\color{black} $_{{\rm 
s}{\rm l}{\rm o}{\rm w}}${\color{black} ) exp(-($\lambda 
$}{\color{black} $_{{\rm t}{\rm r}{\rm a}{\rm n}{\rm 
s}}${\color{black}} {\color{black} 
$t$}{\color{black} )}{\color{black} $^{{\rm \beta 
}}${\color{black} ).}

{\color{black} Common parameters at all temperatures are the 
spectrometer asymmetry} {\color{black} 
$A$}{\color{black} $_{0}${\color{black} = 0.21 and the 
generalised exponents $\alpha $ = 1.15(2) and $\beta $ = 0.64(3) describing 
the shape of the relaxation functions of the slow and fast components, 
respectively. The temperature dependence of the relative asymmetry 
F}{\color{black} $_{{\rm s}{\rm l}{\rm o}{\rm w}}$ of the slow 
component is plotted in Fig. 7, the resulting muon spin relaxation rates 
{\color{black} $\lambda $}{\color{black} $_{{\rm 
l}{\rm o}{\rm n}{\rm g}}$ and {\color{black} $\lambda 
$}{\color{black} $_{{\rm t}{\rm r}{\rm a}{\rm n}{\rm s}}$ in 
Figs. 8 and 9.

{\color{black} The full signal amplitude in the temperature 
regime above 40K reflects a paramagnetic relaxation of the muon spin. An 
additional 10\% signal amplitude increase between 40 K and higher 
temperatures is attributed to a change of the experimental set-up 
(}{\color{black} $i.e.$}{\color{black} , time resolution). 
The increase of the relaxation rate with decreasing temperature (Fig. 8) 
arises from the slowing down of the paramagnetic moments. Between 40 and 12K 
the amplitude of this signal is gradually reduced and a strongly relaxing 
component appears, reflecting a sample volume fraction with quasi-static 
magnetic moments on a time scale of $\sim $10}{\color{black} 
$^{{\rm -} {\rm 7}}${\color{black} s,} {\color{black} 
$e.g.$}{\color{black} inhomogeneous short range magnetic order. The 
ordered volume fraction continuously increases from zero at $\sim $40K to 
100\% at 12K. The $\mu $SR data confirm a magnetic transition between 15 and 
12K, since the low temperature value of F}{\color{black} $_{{\rm 
s}{\rm l}{\rm o}{\rm w}}${\color{black} = 0.29(2) clearly proves 
that below 13K the sample is fully magnetically ordered. The 10\% deviation 
of F}{\color{black} $_{{\rm s}{\rm l}{\rm o}{\rm 
w}}${\color{black} from the theoretical value 1/3 is due to the 
constant sample amplitude fixed to the high temperature value above 40K. The 
reduction of $\lambda $}{\color{black} $_{{\rm l}{\rm o}{\rm 
n}{\rm g }} $  between 40 and 12K reflects the{\color{black}} change 
of the nature of the relaxation mechanism from dynamic in the paramagnetic 
regime above 40K to a quasi-static one of the longitudinal muon spin 
component in the magnetically ordered state. The saturation value of 
{\color{black} $\lambda $}{\color{black} $_{{\rm 
l}{\rm o}{\rm n}{\rm g}} \sim $ 0.27(2)$\mu $s$^{{\rm -} {\rm 1}}$ below 
12K corresponds in a strong collision model to an effective spin fluctuation 
rate of the same order. The very strong transversal relaxation rate 
{\color{black} $\lambda $}{\color{black} $_{{\rm 
t}{\rm r}{\rm a}{\rm n}{\rm s}}$ in the ordered state and the absence of a 
coherent muon spin precession indicate a rather inhomogeneous magnetic state 
with a broad local field distribution at the muon site. The increase of 
{\color{black} $\lambda $}{\color{black} $_{{\rm 
t}{\rm r}{\rm a}{\rm n}{\rm s}}$ with decreasing temperature is compatible 
with the increase of the magnetisation at low temperatures. 
{\color{black} The value at 20K corresponds to a strongly reduced 
signal asymmetry in the intermediate regime between paramagnetic and 
ferromagnetic state and is not directly comparable with those at lower 
temperatures.} 

The fitted value of the shape exponent {\color{black} $\alpha $ 
}is close to 1, the expected value for a quasi-static relaxation of the 
longitudinal signal fraction in the magnetically ordered state. In the 
paramagnetic regime this value is consistent with a homogeneous relaxation 
process with a gaussian local field distribution and a well defined spin 
fluctuation rate. In the ordered state below 13K the exponent 
{\color{black} $\beta $} is significantly smaller than 1, which 
is usually observed in inhomogeneous magnets without spontaneous muon spin 
precession. In Fe$_{{\rm 2}}$VAl the main sources of the static field 
distribution are imperfections of the magnetic lattice caused by diluted 
lattice imperfections and impurities. In this case a lorentzian field 
distribution (exponent $\beta $= 1) is expected for dominant dipolar or 
RKKY-type hyperfine coupling between the muon and the local moments [21]. 
The observed behaviour is an indication for an inhomogeneous magnetic state 
causing a superposition of more than one lorentzian field distribution with 
different static widths. The {\color{black} $\mu $SR} data cannot 
unambiguously clarify the cause for this inhomogeneity. Either size and 
orientation variations of small magnetic moments on each Fe site or diluted 
moments like in a spin glass might be possible. Assuming the latter to be 
the case, we can compare the low temperature saturation value of 
{\color{black} $\lambda $}{\color{black} $_{{\rm 
t}{\rm r}{\rm a}{\rm n}{\rm s}{\rm} }\sim $ 47(3)$\mu $s$^{{\rm -} {\rm 
1}}$ to that of a canonical spin glass like (Au)Fe [22], where a linear 
scaling of \quad {\color{black} $\lambda $}{\color{black} 
$_{{\rm t}{\rm r}{\rm a}{\rm n}{\rm s}}${\color{black} \quad with the 
impurity spin concentration has been observed. From this comparison we 
obtain an impurity spin concentration of about 8(2)\% in 
sc-Fe}{\color{black} $_{{\rm 2}}${\color{black} VAl. 
This value is consistent with the M\"{o}ssbauer results presented below but 
incompatible with the upper bound for Fe II/V site exchange from neutron 
scattering.}  

{\color{black} \textbf{d.) 
M\"{o}ssbauer-spectroscopy:}}{\color{black} 
}{\color{black} $^{{\rm 5}{\rm 7}}${\color{black} 
Fe-M\"{o}ssbauer-spectroscopy experiments were executed on all samples 
Fe}{\color{black} $_{{\rm 2}{\rm +} x}${\color{black} 
V}{\color{black} $_{{\rm 1}{\rm -} x}${\color{black} 
Al, 0 $ \le  x  \le $0.5, in a standard low-temperature M\"{o}ssbauer set-up 
(source:} {\color{black} $^{{\rm 5}{\rm 
7}}${\color{black} Co in Rh matrix) at temperatures ranging from 
8 to 300K. A full account of our experiments will be presented elsewhere 
[23]; here, we only address two aspects of our M\"{o}ssbauer studies on 
Fe}{\color{black} $_{{\rm 2}{\rm +} x}${\color{black} 
V}{\color{black} $_{{\rm 1}{\rm -} x}${\color{black} 
Al.} 

{\color{black} In Fig. 10 we present the M\"{o}ssbauer spectra of 
Fe}{\color{black} $_{{\rm 2}{\rm .}{\rm 
1}}${\color{black} V}{\color{black} $_{{\rm 0}{\rm 
.}{\rm 9}}${\color{black} Al measured at 50 and 10K, together 
with the corresponding spectra of sc-Fe}{\color{black} $_{{\rm 
2}}${\color{black} VAl at 50 and 8K. The experiments at 50K probe 
the paramagnetic state, while at 8/10K the systems are ferromagnetically 
ordered. For both compounds similar M\"{o}ssbauer spectra are observed: at 
50K we detect single lines with small isomer shifts of about 0.02-0.03mm/s, 
relative to} {\color{black} $^{{\rm 5}{\rm 
7}}${\color{black} Co(Rh), and linewidths (FWHM) $\Gamma $ 
0.22mm/s (solid lines in Fig. 10). Upon lowering the temperature below 
}{\color{black} $T$}{\color{black} 
$_{C}${\color{black} a further broadening of the lines and a 
reduction of their depth occurs. The close quantitative and qualitative 
similarity between the temperature dependence of the M\"{o}ssbauer spectra 
of Fe}{\color{black} $_{{\rm 2}{\rm .}{\rm 
1}}${\color{black} V}{\color{black} $_{{\rm 0}{\rm 
.}{\rm 9}}${\color{black} Al and sc-Fe}{\color{black} 
$_{{\rm 2}}${\color{black} VAl is emphasised by plotting the 
difference between the spectra at 50 and 8/10K for both compounds (Fig. 10). 
It reinforces the notion of ferromagnetism in 
sc-Fe}{\color{black} $_{{\rm 2}}${\color{black} VAl as 
result of the same mechanism as in Fe}{\color{black} $_{{\rm 
2}{\rm .}{\rm 1}}${\color{black} V}{\color{black} 
$_{{\rm 0}{\rm .}{\rm 9}}${\color{black} Al and is consistent 
with our view that it arises from negative chemical pressure.} 

{\color{black} For both Fe}{\color{black} $_{{\rm 
2}{\rm .}{\rm 1}}${\color{black} V}{\color{black} 
$_{{\rm 0}{\rm .}{\rm 9}}${\color{black} Al and 
sc-Fe}{\color{black} $_{{\rm 2}}${\color{black} VAl 
the observed behaviour is not the expected one for an archetypical bulk 
ferromagnet, since no well-defined Zeeman splitting is visible. Previously, 
the M\"{o}ssbauer-spectra of Fe}{\color{black} $_{{\rm 2}{\rm + 
}x}${\color{black} V}{\color{black} $_{{\rm 1}{\rm - 
}x}${\color{black} Al have been interpreted in terms of 
fluctuating spins [14,15] and on basis of a ''shell model'' of magnetic ions 
immersed in a non-magnetic matrix [9]. The first model is inconsistent with 
the observation of bulk ferromagnetic ordering for 
Fe}{\color{black} $_{{\rm 2}{\rm .}{\rm 
1}}${\color{black} V}{\color{black} $_{{\rm 0}{\rm 
.}{\rm 9}}${\color{black} Al and sc-Fe}{\color{black} 
$_{{\rm 2}}${\color{black} VAl.} 

{\color{black} The shell model has been successfully applied to 
disordered metallic ferromagnets like Fe-Al alloys [24]. It assumes that the 
hyperfine field at a given Fe ion results from a superposition of the 
contributions of magnetic ions sited in the nearest neighbour, next-nearest 
neighbour etc. shell around the Fe ion. In Fe}{\color{black} 
$_{{\rm 2}{\rm +} x}${\color{black} V}{\color{black} 
$_{{\rm 1}{\rm -} x}${\color{black} Al the magnetic ions are 
thought to be Fe II on V sites, which on basis of band structure 
calculations are predicted to carry a large (2$\mu 
$}{\color{black} $_{{\rm B}}${\color{black} ) magnetic 
moment [6-8]. From our M\"{o}ssbauer data we can estimate for 
sc-Fe}{\color{black} $_{{\rm 2}}${\color{black} VAl 
the required number of site exchanged Fe II/V pairs: to account for the 
broadening of the M\"{o}ssbauer-line below} {\color{black} 
$T$}{\color{black} $_{C}${\color{black} \quad  it would require 
at least 6\% Fe II on V sites. This value, which is more than one order of 
magnitude larger than t}he estimated level of Fe/V site exchanges for 
q-Fe$_{{\rm 2}}$VAl of 0.3-0.4\% [10,11],  {\color{black}  is 
inconsistent with the result of our neutron diffraction experiments, setting 
an upper limit of 3\% Fe II on V sites}. {\color{black} 
Therefore, we conclude that neither model proposed so far properly accounts 
for the M\"{o}ssbauer spectra of Fe}{\color{black} $_{{\rm 2}{\rm 
+} x}${\color{black} V}{\color{black} $_{{\rm 1}{\rm - 
}x}${\color{black} Al,} {\color{black} 
$x$}{\color{black} $ \approx $0.}

\bigskip

\textbf{3.) Conclusions}

We have presented powder neutron diffraction, {\color{black} $\mu 
$SR and M\"{o}ssbauer-experiments on Fe}{\color{black} $_{{\rm 
2}{\rm +} x}${\color{black} V}{\color{black} $_{{\rm 
1}{\rm -} x}${\color{black} Al, and in particular on differently 
heat treated samples Fe}{\color{black} $_{{\rm 
2}}${\color{black} VAl. Our structural investigation proves that 
slowly-cooled Fe}{\color{black} $_{{\rm 
2}}${\color{black} VAl, which we established on a microscopic 
scale to be ferromagnetically ordered below} {\color{black} 
$T$}{\color{black} $_{C}${\color{black} = 13K, is 
structurally better ordered than quenched material, for which we found 
evidence for substantial Al off-stoichiometry. For the quenched material the 
bulk properties [13] resemble the behaviour of ''Kondo-insulating like'' or 
''pseudogap'' Fe}{\color{black} $_{{\rm 
2}}${\color{black} VAl [1,2,5,9-11,16]; this suggests that the 
materials investigated in those works are Al deficient.} 

{\color{black} As pointed out in Ref. [5], because of the 
electron count Al deficiency might have a strong effect on the actual 
position of the Fermi level} {\color{black} 
$E$}{\color{black} $_{F}${\color{black} in a pseudogap 
system. Specifically, it was argued that Al deficiency moves 
}{\color{black} $E$}{\color{black} 
$_{F}${\color{black} out of the centre of the pseudogap into the 
slopes, implying that Al deficient material should be more metallic than 
non-deficient one. This hypothesis is in conflict with the observation that 
the resistivity of sc-Fe}{\color{black} $_{{\rm 
2}}${\color{black} VAl is lower than that of 
q-Fe}{\color{black} $_{{\rm 2}}${\color{black} VAl, 
indicating less metallicity and} {\color{black} 
$E$}{\color{black} $_{F}${\color{black} in the centre of 
the gap for the latter sample.} 

{\color{black} Another explanation for the smaller resistivity of 
sc-Fe}{\color{black} $_{{\rm 2}}${\color{black} VAl 
might be based upon the ability of Al vacancies to localise conduction 
electrons. Then, the resistivity of Fe}{\color{black} $_{{\rm 
2}}${\color{black} VAl would depend strongly on disorder induced 
(}{\color{black} $i.e.$}{\color{black} , from Al vacancies) 
localisation of electrons in states close to the gap; with the larger degree 
of disorder in q-Fe}{\color{black} $_{{\rm 
2}}${\color{black} VAl its resistivity will be larger than that 
of sc-Fe}{\color{black} $_{{\rm 2}}${\color{black} 
VAl. Unfortunately, quantitative predictions are very difficult for this 
scenario, as it depends both on the actual position of 
}{\color{black} $E$}{\color{black} 
$_{F}${\color{black} because of the number of charge carriers 
from Al, and the strength and number of localising potentials from Al 
vacancies. In addition,} {\color{black} 
$E$}{\color{black} $_{F}${\color{black} might depend on 
the lattice parameter,} {\color{black} 
$i.e.$}{\color{black} , the chemical pressure, which in turn is a 
function of the Al stoichiometry.}

{\color{black} Altogether, we conclude} that the physical 
properties, in particular the anomalous resistivity, of non-magnetic 
Fe$_{{\rm 2}}$VAl are dominated by crystallographic disorder. 
Kondo-insulating behaviour seems not to play a role; specifically, Fe$_{{\rm 
2}}$VAl can be easily tuned into a ferromagnet, which is not expected for 
inherently non-magnetic Kondo insulators [25]. Further, we conclude that 
sc-Fe$_{{\rm 2}}$VAl is ''more representative'' of the intrinsic behaviour 
of ambient pressure Fe$_{{\rm 2}}$VAl than quenched material, as it is 
closer to perfect stoichiometry. Hence, while the closeness to a pseudogap 
state predicted in band structure calculations [6-8] is found for both sc- 
and q-Fe$_{{\rm 2}}$VAl, the prediction of a non-magnetic ground state for 
stoichiometric, perfectly Heusler-ordered Fe$_{{\rm 2}}$VAl must still be 
considered a matter of debate. {\color{black} In particular, we 
note that, given our M\"{o}ssbauer- and $\mu $SR-experiments, the simple 
view of ''Fe on the wrong (V) site'' generating magnetism does not account 
for magnetic order in} sc-Fe$_{{\rm 2}}$VAl, in so far as the measured 
number of ''wrong sited'' Fe is too low. 

Since the properties of Fe$_{{\rm 2}}$VAl are so sensitively dependent on 
the actual stoichiometry, which is hard to control in the limits relevant to 
Fe$_{{\rm 2}}$VAl, we believe that pressure experiments on either 
sc-Fe$_{{\rm 2}}$VAl or q-Fe$_{{\rm 2}}$VAl represent a more fruitful route 
to study this material. Based on our experiments, for sc-Fe$_{{\rm 2}}$VAl 
we would expect a suppression of magnetic order upon application of 
pressure. For q-Fe$_{{\rm 2}}$VAl a pressure experiment might yield insight 
in the relevance of Al vacancies for electronic localisation and the 
position of the Fermi level. Furthermore, pressure experiments might be very 
useful in combination with band structure calculations, as we would expect 
that the result of these calculations should sensitively depend on the value 
of the lattice constant.

\bigskip

\textbf{Acknowledgements:} 

{\color{black} Work at the TU Braunschweig was supported by the 
Deutsche Forschungsgemeinschaft DFG, under Grant No. SU 229/4-1.}

\bigskip

{\color{black} \underline {\textbf{References:}}}

{\color{black} [1] Nishino Y, Kato M, Asano S, Soda K, Hayasaki 
M, Mizutani U 1997} {\color{black} \textit{Phys. Rev. Lett.}}{\color{black} 
}{\color{black} \textbf{ 79 }}{\color{black}   1909}

{\color{black} [2] Endo K, Matsuda H, Ooiwa K, Iijima M, Goto T, 
Sato K, Umehara I 1998} {\color{black} \textit{J. Magn. Magn. Mat.}}{\color{black} 
}{\color{black} \textbf{ 177-181 }}{\color{black} 1437}

{\color{black} [3] Schlesinger Z, Fisk Z, Zhang H-T, Maple M B, 
DiTusa J F, Aeppli G 1993} {\color{black} \textit{Phys. Rev. Lett.}}{\color{black} 
}{\color{black} \textbf{ 71 }}{\color{black} 1748;} 

{\color{black} [4] Schlesinger Z, Fisk Z, Zhang H-T, Maple M B 
1997} {\color{black} \textit{Physica B}}{\color{black} 
}{\color{black} \textbf{ 237-38 }}{\color{black} 460}

{\color{black} [5] Okamura H, Kawahara J, Namba T, Kimura S, Soda 
K, Mizutani U, Nishino Y, Kato M, Shimoyama I, Miura H, Fukui K, Nakagawa K, 
Nakagawa H, Kinoshita T 2000} {\color{black} \textit{Phys. Rev. Lett.}}{\color{black} 
}{\color{black} \textbf{ 84 }}{\color{black} 3674}

{\color{black} [6] Guo G Y, Botton G A, Nishino Y 1998 
}{\color{black} \textit{J. Phys.: Condens. Matter}}{\color{black} 
}{\color{black} \textbf{ 10 }}{\color{black} L119}

{\color{black} [7] Singh D J and Mazin I I, 1998 
}{\color{black} \textit{Phys. Rev. B}}{\color{black} 
}{\color{black} \textbf{ 57 }}{\color{black} 14352}

{\color{black} [8] Weht R and Pickett W E 1998 
}{\color{black} \textit{Phys. Rev. B}}{\color{black} 
}{\color{black} \textbf{ 58 }}{\color{black} 6855}

{\color{black} [9] Matsuda H, Endo K, Ooiwa K, Iijima M, Takano 
Y, Mitamura H, Goto T, Tokiyama M, Arai J 2000} {\color{black} 
\textit{J. Phys. Soc. Jpn.}}{\color{black}} {\color{black} 
\textbf{ 69 }}{\color{black} 1004}

{\color{black} [10] Lue C-S and Ross Jr. J H 1998 
}{\color{black} \textit{Phys. Rev. B}}{\color{black} 
}{\color{black} \textbf{ 58 }}{\color{black} 9763}

{\color{black} [11] Lue C-S, Ross Jr. J H, Chang C F, Yang H D 
1999} {\color{black} \textit{Phys. Rev. B}}{\color{black} 
}{\color{black} \textbf{ 60 }}{\color{black} R13941}

{\color{black} [12] Slebarski A, Maple M B, Freeman E J, Sirvent 
C, Tworuszka C, Orzechowska M, Wrona A, Jezierski A, Chiuzbaian S, Neumann M 
2000} {\color{black} \textit{Phys. Rev. B}}{\color{black} 
}{\color{black} \textbf{ 62 }}{\color{black} 3296}

{\color{black} [13] Matsushita A and Yamada Y 1999 
}{\color{black} \textit{J. Magn. Magn. Mat.}}{\color{black} 
}{\color{black} \textbf{ 196-197 }}{\color{black} 669}

{\color{black} [14] Popiel E S, Tuszynski M, Zarek W, Rendecki T 
1989} {\color{black} \textit{J. Less-Common Met.} }{\color{black} \textbf{ 146  
}}{\color{black} 127}

{\color{black} [15] Popiel E S, Zarek W, Tuszynski M 1989 
}{\color{black} \textit{Hyperfine Interactions}}{\color{black} 
}{\color{black} \textbf{ 51 }}{\color{black} 981}

[16] Kato M, Nishino Y, Mizutani U, Asano S 2000 \textit{J. Phys.: Condens. Matter} \textbf{ 12 } 1769

[17] T\"{o}bbens D M, St\"{u}{\ss}er N, Knorr K, Mayer H M, Lampert G, Proc. 
of the 7. European Powder Diffraction Conference EPDIC 2000

[18] Rodriguez-Carvajal J, Laboratoire Leon Brillouin -- CEA-CNRS, Version 
3.5d Oct98-LLB-JRC

{\color{black} [19] Nachumi B, Fudamoto Y, Keren A, Kojima K M, 
Larkin M, Luke G M, Merrin J, Tchernyshyov O, Uemura Y J, Ichikawa N, Goto 
M, Takagi H, Uchida S, Crawford M K, McCarron E M, MacLaughlin D E, Heffner 
R H 1998} {\color{black} \textit{Phys. Rev. B}}{\color{black} 
}{\color{black} \textbf{ 58 }}{\color{black} 8760}

{\color{black} [20] Dalmas de R\'{e}otier P and Yaouanc A 1997 
}{\color{black} \textit{J. Phys.: Condens. Matter}}{\color{black} 
}{\color{black} \textbf{ 9 }}{\color{black} 9113}

[21] Walstedt R E and Walker L P 1974 {\color{black} \textit{Phys. Rev. B}}{\color{black} 
}{\color{black} \textbf{ 9 }}{\color{black} 4857}

[22] Uemura Y J and Yamazaki T 1983 J. Mag. Mag. Mat \textbf{ 31-34 } 1359

{\color{black} [23] Baabe D} {\color{black} \textit{et al.}}{\color{black} 
, in preparation} 

[24] Niculescu V, Raj K, Budnick J L, Burch T J, Hines W A, and Menotti AH 
1976 \textit{Phys. Rev. B} \textbf{ 14 } 4160

[25] Aeppli G and Fisk Z 1992 \textit{Comment Cond. Mat. Phys.} \textbf{ 16 } 155

\bigskip

{\color{black} {\textbf{Figure captions}}}

\bigskip

Fig. 1: a.) The cubic \textit{DO}$_{3}$ lattice of Fe$_{{\rm 3}}$Al, with the two 
inequivalent Fe sites at Fe I: ($\raise.5ex\hbox{$\scriptstyle 
1$}\kern-.1em/ \kern-.15em\lower.25ex\hbox{$\scriptstyle 2$} $ 
$\raise.5ex\hbox{$\scriptstyle 1$}\kern-.1em/ 
\kern-.15em\lower.25ex\hbox{$\scriptstyle 2$} $ 
$\raise.5ex\hbox{$\scriptstyle 1$}\kern-.1em/ 
\kern-.15em\lower.25ex\hbox{$\scriptstyle 2$} $) and Fe II: 
($\raise.5ex\hbox{$\scriptstyle 1$}\kern-.1em/ 
\kern-.15em\lower.25ex\hbox{$\scriptstyle 4$} $ 
$\raise.5ex\hbox{$\scriptstyle 1$}\kern-.1em/ 
\kern-.15em\lower.25ex\hbox{$\scriptstyle 4$} $ 
$\raise.5ex\hbox{$\scriptstyle 1$}\kern-.1em/ 
\kern-.15em\lower.25ex\hbox{$\scriptstyle 4$} 
$);($\raise.5ex\hbox{$\scriptstyle 3$}\kern-.1em/ 
\kern-.15em\lower.25ex\hbox{$\scriptstyle 3$} $ 
$\raise.5ex\hbox{$\scriptstyle 3$}\kern-.1em/ 
\kern-.15em\lower.25ex\hbox{$\scriptstyle 3$} $ 
$\raise.5ex\hbox{$\scriptstyle 3$}\kern-.1em/ 
\kern-.15em\lower.25ex\hbox{$\scriptstyle 3$} $). b.) The calculated 
normalised neutron diffraction intensities for the three different 
structural modifications Fe$_{{\rm 2}}$VAl: the fully ordered Heusler 
lattice (V on Fe I site), 10\% Fe II/V site exchange and V entirely 
occupying one Fe II site, giving rise to a superstructure. 

\bigskip

Fig. 2: The neutron diffraction spectrum of Fe$_{{\rm 2}{\rm .}{\rm 
1}}$V$_{{\rm 0}{\rm .}{\rm 9}}$Al, measured at 50K (+). Included is a 
refined fit to the data, the difference between fit and data, and tics 
indicating Bragg peak positions.

\bigskip

Fig. 3: The lattice parameters of different samples Fe$_{{\rm 2}{\rm + 
}x}$V$_{{\rm 1}{\rm -} x}$Al, 0 {\color{black} $ \le $} $x$ 
{\color{black} $ \le $} 0.1, as function of $x$. The inset depicts 
the data for 0 $ \le  x  \le $.5. Lines are guides to the eye.

\bigskip

Fig. 4: The normalised neutron spectra of sc- (solid line) and q-Fe$_{{\rm 
2}}$VAl (+), and the difference I$_{{\rm s}{\rm c}{\rm -} {\rm F}{\rm e}{\rm 
2}{\rm V}{\rm A}{\rm l}}$-I$_{{\rm q}{\rm -} {\rm F}{\rm e}{\rm 2}{\rm 
V}{\rm A}{\rm l}}$ between the two spectra.

\bigskip

Fig. 5: The relative Bragg peak intensities of sc-Fe$_{{\rm 2}}$VAl and the 
intensity variations, calculated from 1 - $\Sigma $ (I$_{{\rm q}{\rm -} {\rm 
F}{\rm e}{\rm 2}{\rm V}{\rm A}{\rm l}}$ ($\Theta $) / I$_{{\rm s}{\rm c}{\rm 
-} {\rm F}{\rm e}{\rm 2}{\rm V}{\rm A}{\rm l}}$($\Theta $))$\Delta \Theta 
$.

\bigskip

{\color{black} Fig. 6: Zero field $\mu $SR spectra of 
sc-Fe}{\color{black} $_{{\rm 2}}${\color{black} VAl at 
different temperatures. The fit functions (solid lines) are described in the 
text.}

\bigskip

{\color{black} Fig. 7: Temperature dependence of the slowly 
relaxing signal fraction F}{\color{black} $_{{\rm s}{\rm l}{\rm 
o}{\rm w}{\rm} }${\color{black} in zero field $\mu $SR of 
sc-Fe}{\color{black} $_{{\rm 2}}${\color{black} VAl.}

\bigskip

{\color{black} Fig. 8: Temperature dependence of the longitudinal 
muon spin relaxation rate $\lambda $}{\color{black} $_{{\rm 
l}{\rm o}{\rm n}{\rm g}} ${\color{black} in zero field $\mu $SR 
of sc-Fe}{\color{black} $_{{\rm 2}}${\color{black} 
VAl.}

\bigskip

{\color{black} Fig. 9: Temperature dependence of the transversal 
muon spin relaxation rate $\lambda $}{\color{black} $_{{\rm 
t}{\rm r}{\rm a}{\rm n}{\rm s}} ${\color{black} in zero field 
$\mu $SR of sc-Fe}{\color{black} $_{{\rm 
2}}${\color{black} VAl.}

\bigskip

{\color{black} Fig. 10: M\"{o}ssbauer-transmission-spectra of 
Fe}{\color{black} $_{{\rm 2}{\rm .}{\rm 
1}}${\color{black} V}{\color{black} $_{{\rm 0}{\rm 
.}{\rm 9}}${\color{black} Al and sc-Fe}{\color{black} 
$_{{\rm 2}}${\color{black} VAl taken at 50 (+) and 10/8K ($o$. 
Solid lines indicate single lorentzian fits to the 50K data. Included in the 
plot are the differences between the spectra at 50 and 10/8K ($o$, 
respectively.}

\bigskip

{\color{black} \underline {\textbf{Tables}}}

\newcommand{\PreserveBackslash}[1]{\let\temp=\\#1\let\\=\temp}
\let\PBS=\PreserveBackslash
\begin{table}
\begin{tabular}
{|p{67pt}|p{61pt}|p{50pt}|p{47pt}|p{90pt}|p{40pt}|}
\hline
Sample& 
Lattice \par parameter $a$ ({\AA})& 
B$_{{\rm i}{\rm s}{\rm o}{\rm -} {\rm F}{\rm e}}$ ({\AA}$^{{\rm 2}}$)& 
B$_{{\rm i}{\rm s}{\rm o}{\rm -} {\rm V}{\rm /} {\rm A}{\rm l}}$ ({\AA}$^{{\rm 2}}$)& 
Measured composition& 
R$_{{\rm B}{\rm r}{\rm a}{\rm g}{\rm g}}$ \\
\hline
Fe$_{{\rm 2}{\rm .}{\rm 5}}$V$_{{\rm 0}{\rm .}{\rm 5}}$Al& 
5.7604(3)& 
0.15(8)& 
0.9(2)& 
Fe$_{{\rm 2}{\rm .}{\rm 5}}$V$_{{\rm 0}{\rm .}{\rm 5}}$Al& 
4.3 \%  \\
\hline
Fe$_{{\rm 2}{\rm .}{\rm 1}}$V$_{{\rm 0}{\rm .}{\rm 9}}$Al& 
5.7528(2)& 
0.15(6)& 
0.3(2)& 
Fe$_{{\rm 2}{\rm .}{\rm 1}}$V$_{{\rm 0}{\rm .}{\rm 9}}$Al$_{{\rm 1}{\rm .}{\rm 0}{\rm 1}{\rm (}{\rm 5}{\rm )}}$& 
2.0 \%  \\
\hline
Fe$_{{\rm 2}{\rm .}{\rm 0}{\rm 2}}$V$_{{\rm 0}{\rm .}{\rm 9}{\rm 8}}$Al& 
5.7521(2)& 
0.13(6)& 
0.4(2)& 
Fe$_{{\rm 2}{\rm .}{\rm 0}{\rm 2}}$V$_{{\rm 0}{\rm .}{\rm 9}{\rm 8}}$Al$_{{\rm 1}{\rm .}{\rm 0}{\rm 1}{\rm (}{\rm 4}{\rm )}}$& 
2.1 \%  \\
\hline
sc-Fe$_{{\rm 2}}$VAl& 
5.7523(2)& 
0.25(6)& 
0.4(3)& 
Fe$_{{\rm 2}}$VAl$_{{\rm 0}{\rm .}{\rm 9}{\rm 9}{\rm (}{\rm 4}{\rm )}}$& 
1.9 \%  \\
\hline
q-Fe$_{{\rm 2}}$VAl & 
5.7514(2)& 
0.23(6)& 
0.3(3)& 
Fe$_{{\rm 2}}$VAl$_{{\rm 0}{\rm .}{\rm 9}{\rm 4}{\rm (}{\rm 4}{\rm )}}$& 
2.3 \%  \\
\hline
\end{tabular}
\end{table}

Table 1: Summary of the refinement results of the neutron diffraction data 
on Fe$_{{\rm 2}{\rm +} x}$V$_{{\rm 1}{\rm -} x}$Al, 
{\color{black} 0 $ \le $} {\color{black} 
$x$}{\color{black} $ \le $ 0.5}, with the cubic lattice parameter 
$a$, the isotropic thermal displacement parameters B$_{{\rm i}{\rm s}{\rm o}}$ 
of Fe and V/Al, the measured composition and the value of R$_{{\rm B}{\rm 
r}{\rm a}{\rm g}{\rm g}}$.$_{{\rm} }$

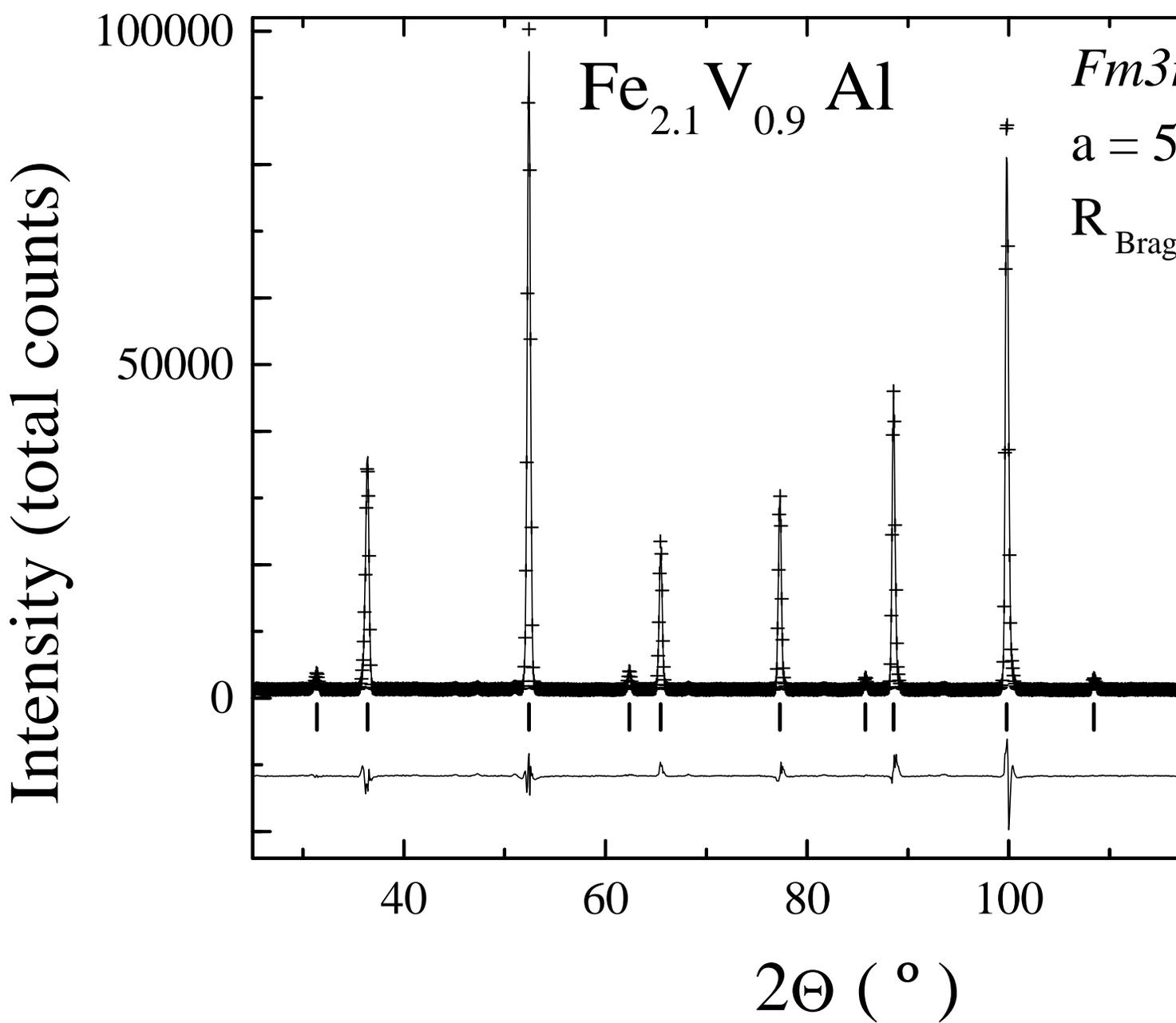
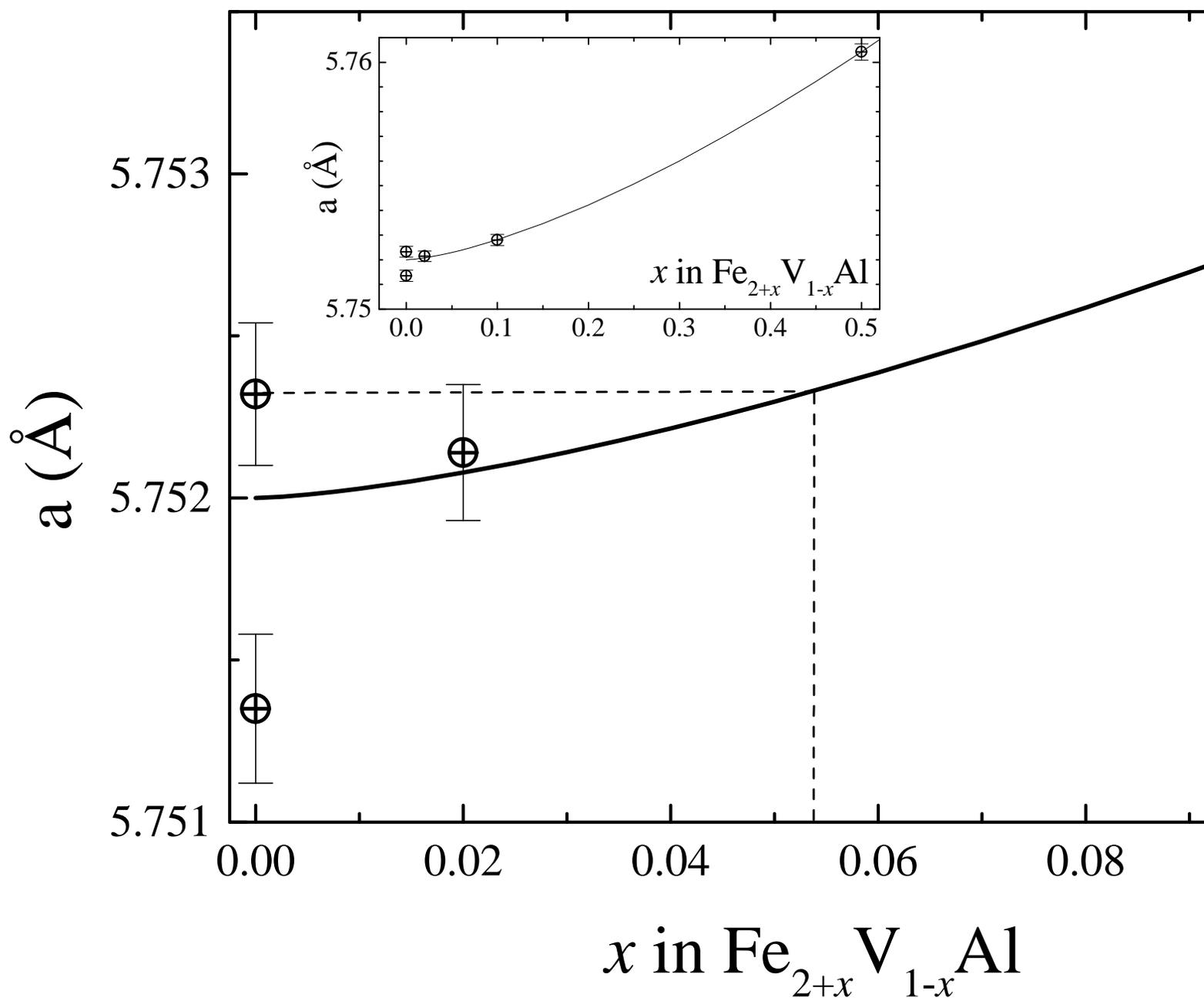
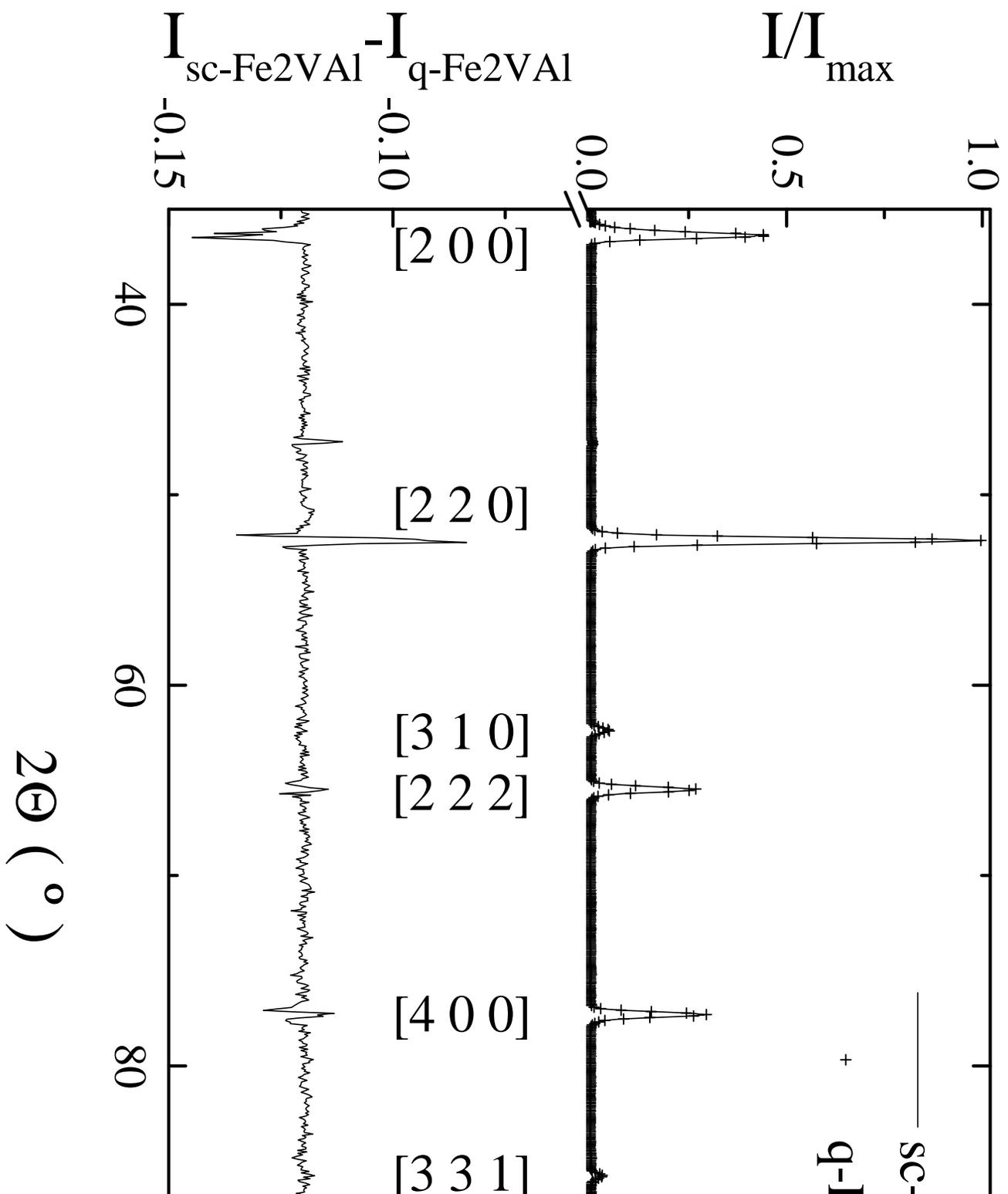
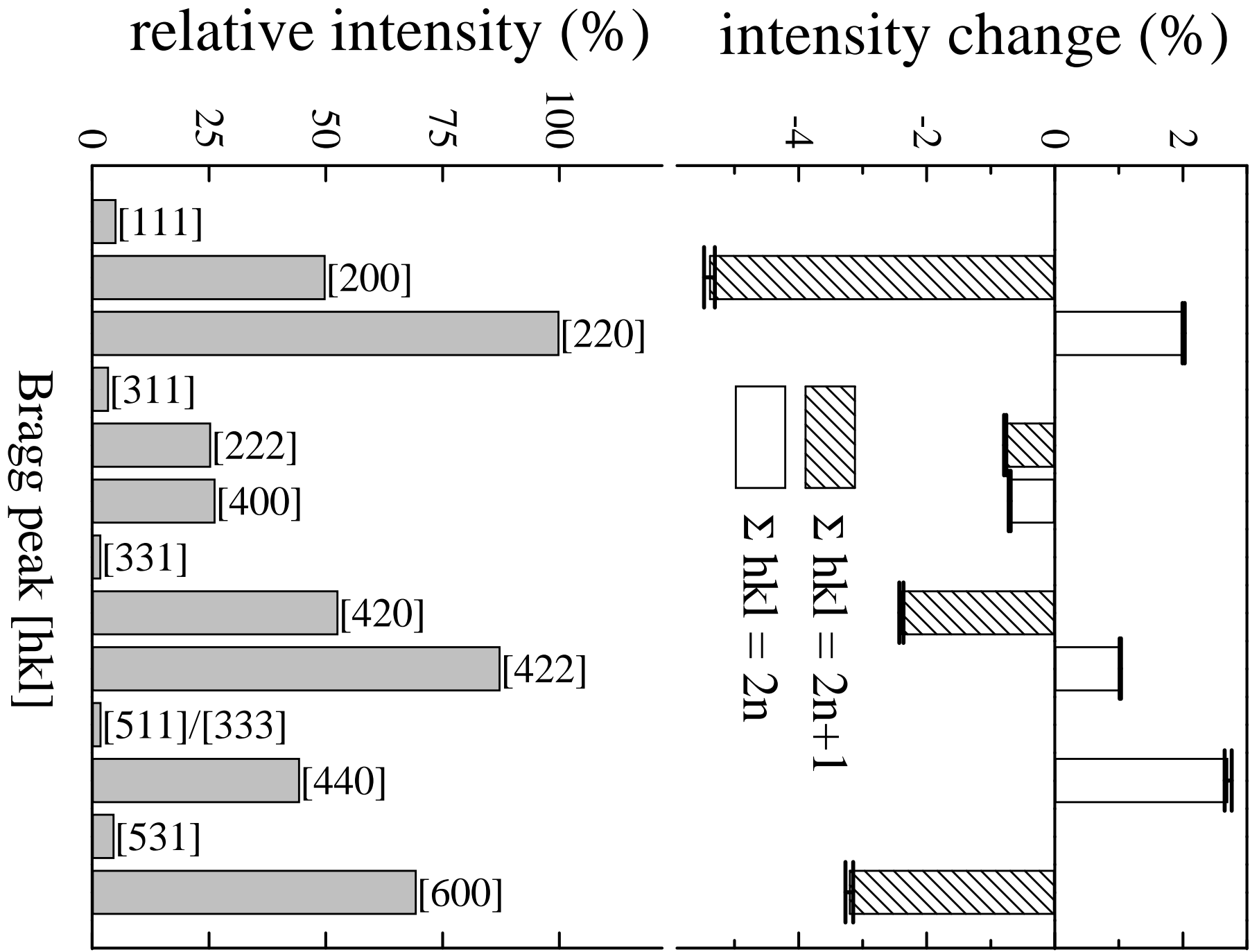
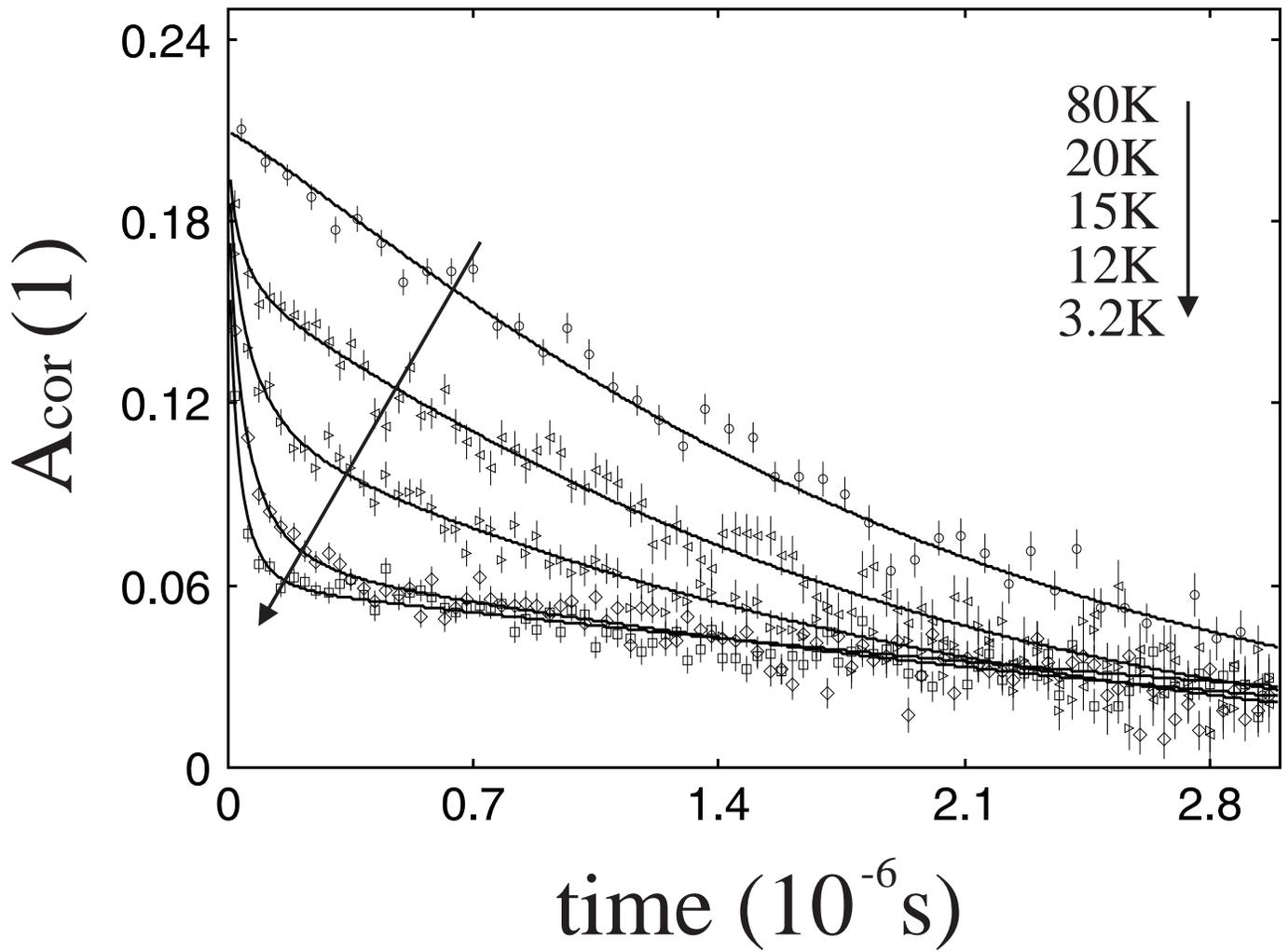
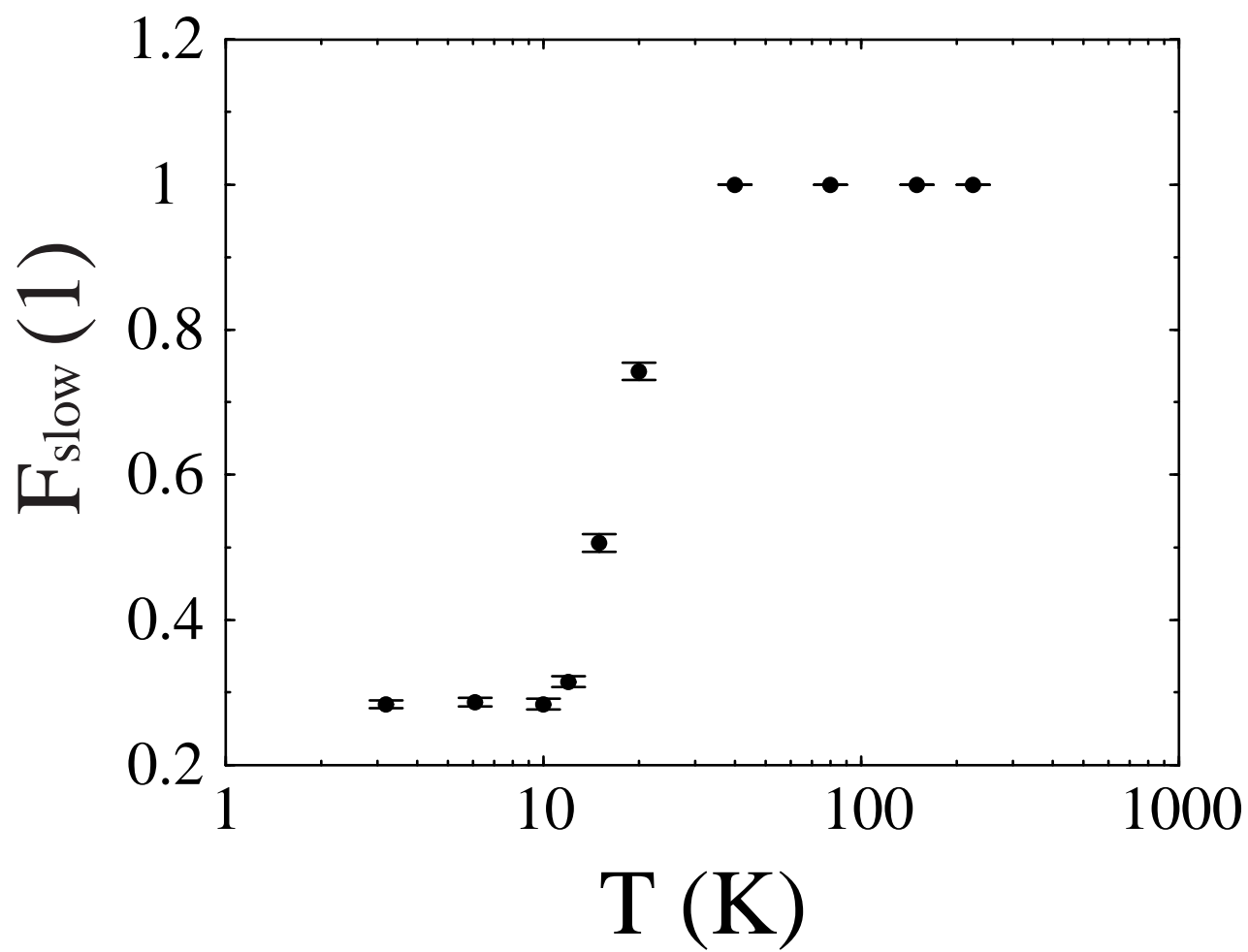
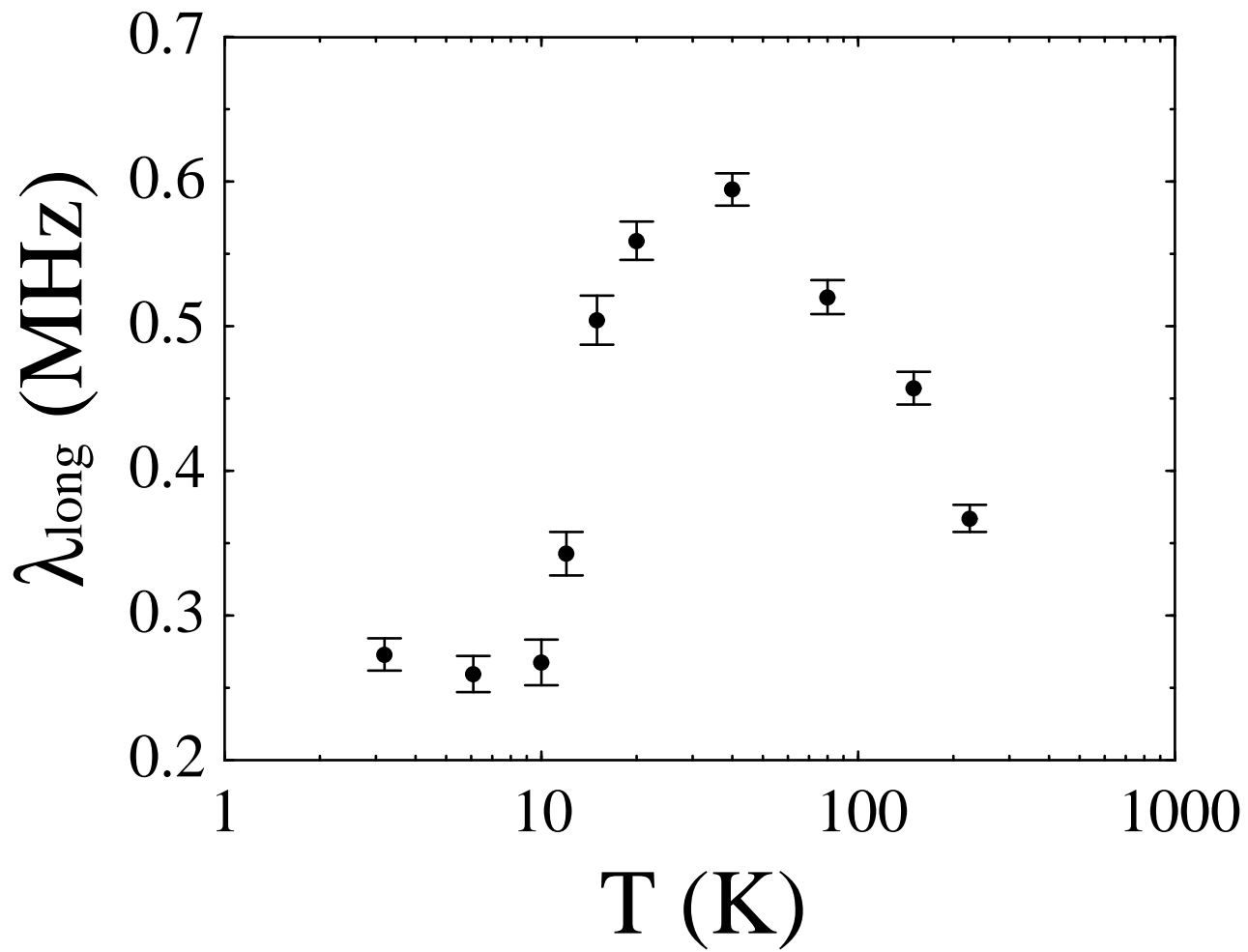
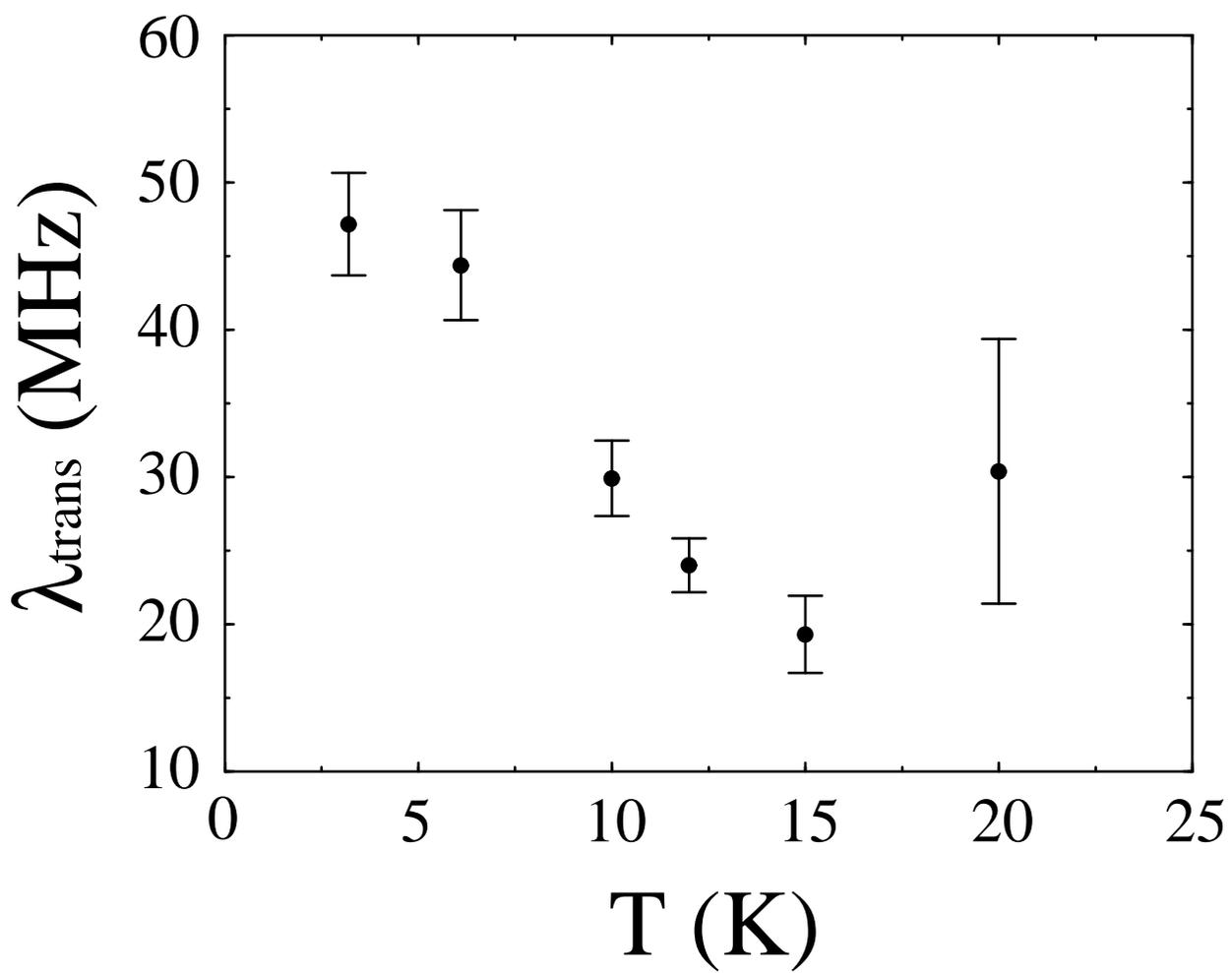
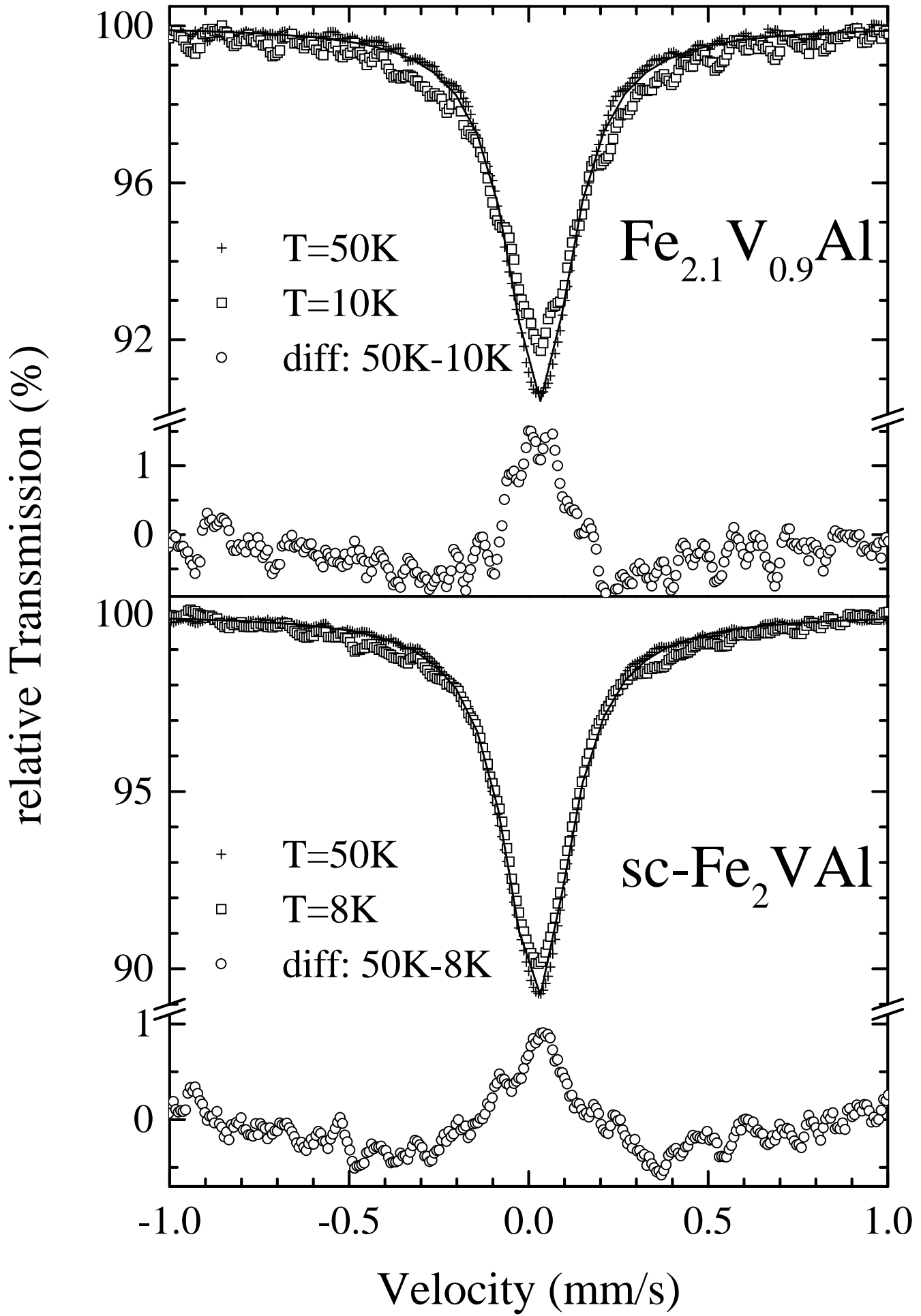
\end{document}